\documentclass[journal]{IEEEtran}

\usepackage{amsmath,amssymb,amsfonts}
\usepackage[utf8]{inputenc}
\usepackage{algorithm}
\usepackage[noend]{algpseudocode}
\algrenewcommand\algorithmicrequire{\textbf{Input:}}
\algrenewcommand\algorithmicensure{\textbf{Output:}}
\usepackage{array}
\usepackage{textcomp}
\usepackage{url}
\usepackage{graphics}
\usepackage{graphicx}
\usepackage{verbatim}
\usepackage{cite} 
\usepackage{siunitx}
\usepackage{booktabs, makecell}
\ifCLASSOPTIONcompsoc
    \usepackage[caption=false, font=normalsize, labelfont=sf, textfont=sf]{subfig}
\else
    \usepackage[caption=false, font=footnotesize]{subfig}
\fi
\usepackage{nomencl}
\usepackage{multirow,tabularx}
\usepackage{etoolbox}
\usepackage{array,multirow}
\usepackage{colortbl}
\usepackage{mathtools}
\usepackage{mdwtab}
\makeatletter
\newcommand*{\rom}[1]{\expandafter\@slowromancap\romannumeral #1@}
\makeatother
\usepackage{float}
\graphicspath{{./Figures/}}
\usepackage{comment}
\usepackage{gensymb} 
\usepackage{bm} 
\usepackage{cleveref} 
\usepackage[table]{xcolor}
\usepackage{diagbox}
\newcolumntype{C}{>{\centering\arraybackslash}X}
\def\BibTeX{{\rm B\kern-.05em{\sc i\kern-.025em b}\kern-.08em
T\kern-.1667em\lower.7ex\hbox{E}\kern-.125emX}}

\begin{document}

\bstctlcite{IEEEexample:BSTcontrol}   

\title{A Novel Power-Band based Data Segmentation Method for Enhancing Meter Phase and Transformer-Meter Pairing Identification}

\author{Han~Pyo~Lee,~\IEEEmembership{Student Member,~IEEE}, PJ~Rehm, Matthew~Makdad, Edmond~Miller, Ning~Lu,~\IEEEmembership{Fellow,~IEEE}
\thanks{This research is supported by the U.S. Department of Energy's Office of Energy Efficiency and Renewable Energy (EERE) under the Solar Energy Technologies Office Award Number DE-EE0008770. Han Pyo Lee and Ning Lu are with the Electrical \& Computer Engineering Department, Future Renewable Energy Delivery and Management (FREEDM) Systems Center, North Carolina State University, Raleigh, NC 27606 USA. (e-mails: hlee39@ncsu.edu, nlu2@ncsu.edu). PJ Rehm is with ElectriCities. Matthew Makdad and Edmond Miller are with New River Light and Power.}
\vspace{-0.3in}}



\maketitle

\begin{abstract}
This paper presents a novel power-band-based data segmentation (PBDS) method to enhance the identification of meter phase and meter-transformer pairing. Meters that share the same transformer or are on the same phase typically exhibit strongly correlated voltage profiles. However, under high power consumption, there can be significant voltage drops along the line connecting a customer to the distribution transformer. These voltage drops significantly decrease the correlations among meters on the same phase or supplied by the same transformer, resulting in high misidentification rates. To address this issue, we propose using power bands to select highly correlated voltage segments for computing correlations, rather than relying solely on correlations computed from the entire voltage waveforms. The algorithm's performance is assessed by conducting tests using data gathered from 13 utility feeders. To ensure the credibility of the identification results, utility engineers conduct field verification for all 13 feeders. The verification results unequivocally demonstrate that the proposed algorithm surpasses existing methods in both accuracy and robustness.
\end{abstract}


\vspace{-0.1in}
\section{Introduction}
\IEEEPARstart{A}{ccurate} meter phase and meter-transformer pairing information is crucial for efficient distribution circuit operation, planning, and asset management in modern utility systems. Manual registration of this data is error-prone, and updates are frequently required for circuit upgrades, reconfiguration, and equipment replacements. However, the large number of distribution transformers and customer meters makes manual verification costly. As a result, utility engineers are actively exploring data-driven, automated methods for periodic information verification.

The increasing adoption of smart meters \cite{guerrero2021data} has provided utilities with time-series power and voltage measurements at sampling rates of 15 or 30 minutes \cite{FERC}. This availability of data has made leveraging smart meter data for automated identification of phase and transformer-meter pairing relationships an appealing option in recent years. Table \ref{tab1} presents a comprehensive overview of existing methods for addressing phase and transformer-meter pairing identification problems, along with a comparison of their strengths and weaknesses in relation to the proposed approach introduced in this paper. 

As shown in the table, in the realm of phase identification problems, there are two main approaches developed: directly measuring methods \cite{shen2013three, wen2015phase} and data-driven methods. With the widespread deployment of smart meters, data-driven methods have become the preferred choice for utilities. Table I provides an overview of the various data-driven approaches, including real power-based \cite{arya2011phase, arya2013inferring}, voltage-based \cite{pezeshki2012consumer, olivier2018phase}, and machine learning-based approaches \cite{mitra2015voltage, wang2016phase, hosseini2020machine, foggo2019improving, blakely2019spectral, blakely2020phase}. Notably, machine learning (ML)-based methods have witnessed a surge in popularity in recent years. Building upon our prior work presented in \cite{lee2022novel}, which introduced a machine-learning based phase identification method, this paper extends the methodology to use real smart meter data as inputs instead of synthetic data. Furthermore, we validate its performance by conducting field verification executed by utility engineers, thereby affirming its superiority over existing methods.

\begin{table*}[ht]
	\begin{center}
		\caption{A Review of Existing Methods and Our Contributions.}
		\vspace{-0.1in}
		\label{tab1}
		\centerline{\includegraphics[width=\linewidth, height=0.35\textheight]{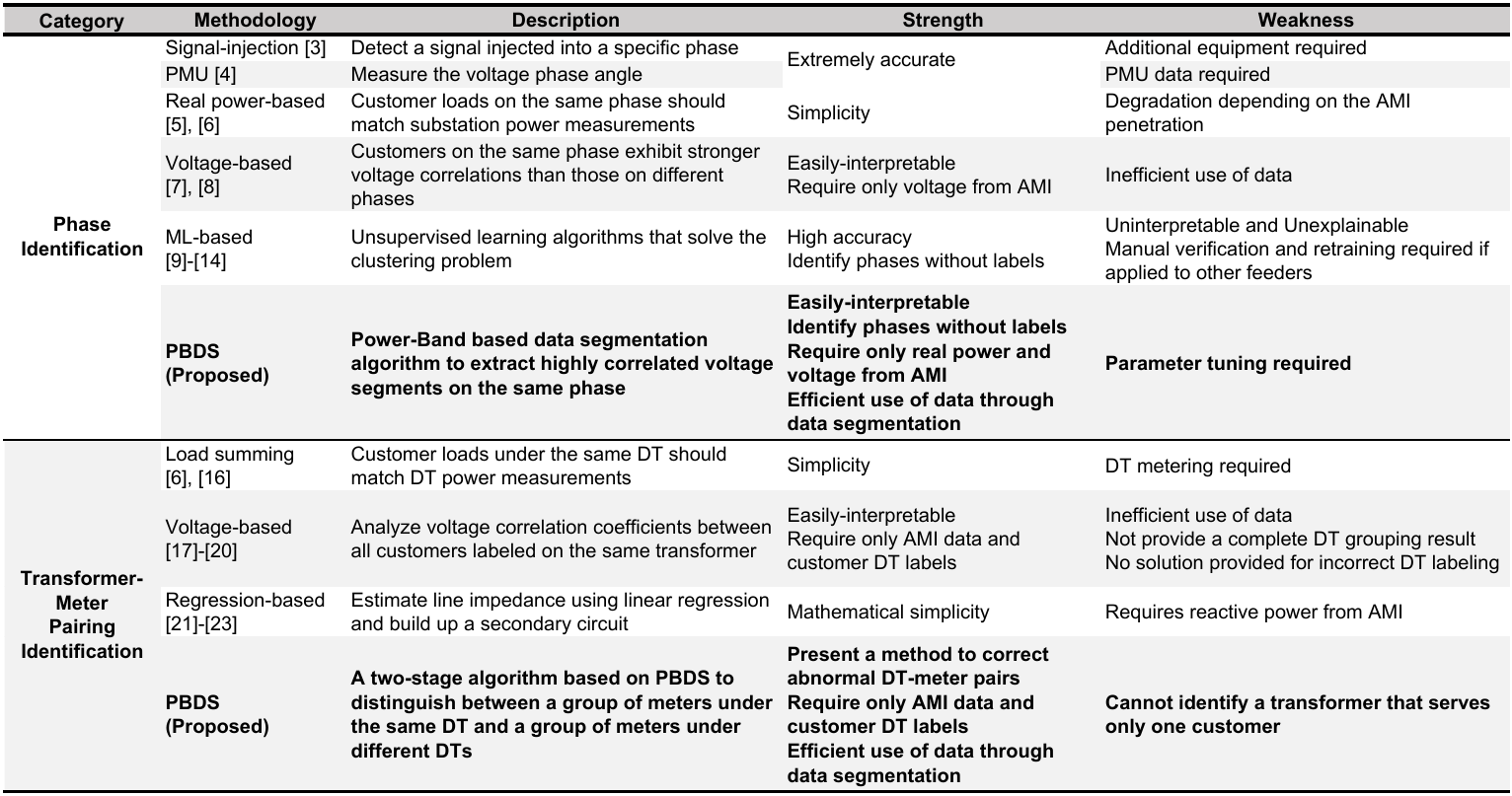}}  
	\end{center}
\vspace{-0.4in}
\end{table*}

In the realm of transformer-meter pairing, there are three main approaches: real power-based, voltage-based, and linear regression-based techniques, also highlighted in Table \ref{tab1}. The process of identifying transformer-meter pairings shares noticeable similarities with meter phase identification. Among the real power-based methods, the load-summing approach \cite{pappu2017identifying, arya2013inferring} has gained significant popularity. Similar to meter phase identification, voltage correlation-based methods are also commonly employed for transformer-meter pairing identification \cite{luan2015smart, watson2016use, weng2016distributed, blakely2020identifying}. In recent years, linear regression-based methods have attracted increasing attention due to their mathematical simplicity in system topology identification \cite{short2012advanced, ye2019two, blakely2021identification}.

Upon comparing the methods, it becomes evident that the calculation of voltage correlation plays a pivotal role in meter topology identification, regardless of whether it pertains to phase identification or pairing identification. However, current state-of-the-art techniques suffer from two significant limitations. Firstly, all existing methods rely on utilizing the entire time-series data for voltage correlation computation. Nevertheless, as demonstrated in Section \ref{section2}, not all segments of the voltage profile exhibit high correlation among meters sharing the same phase or supplied by the same transformer. Incorporating numerous less-correlated voltage segments leads to a deterioration in correlation, resulting in increased false positive rates. Secondly, the majority of algorithms focus solely on voltage measurements as inputs, disregarding the valuable information provided by power measurements. Leveraging this additional data could effectively enhance the accuracy and robustness of voltage-based identification algorithms.

Therefore, in this paper, we present a novel power-band-based data segmentation (PBDS) method that leverages both power and voltage inputs. To establish the foundation for our proposed algorithm, we conduct circuit analysis to illustrate the occurrence of voltage correlation deterioration phenomena and to provide a theoretical framework for the application of the PBDS method. This new approach using power-bands and minimum duration thresholds to select highly correlated voltage data segments from the time-series voltage profiles of a pair of meters to obtain an ensemble of segments. Then, voltage correlations of the two meters can be calculated from the ensemble instead of using two entire voltage profiles. 

The primary contribution of the PBDS approach is the use of power-bands for data segmentation, leading to a substantial improvement in the accuracy of downstream tasks such as clustering phase groups or identifying transformer-meter pairs. The concept of selecting highly correlated voltage segments using power bands, along with a duration threshold for excluding short data segment, is a novel idea that has not yet been explored in existing literature. To evaluate the effectiveness of this approach, extensive testing was conducted on 13 real utility feeders. Most importantly, the performance improvement achieved by our approach, compared to existing methods, has been fully validated through extensive field verification conducted by utility engineers. 

The rest of this paper is organized as follows. Section~\rom{2} introduces the phase and transformer-meter pairing identification problems, the correlation deterioration phenomena, and PBDS principles. Section~\rom{3} presents the application of PBDS in meter phase identification and in transformer-meter pairing identification. Simulation results are presented in Section~\rom{4} and Section~\rom{5} concludes the paper.

\begin{figure}[ht]
\centerline{\includegraphics[width=\linewidth, height=0.18\textheight]{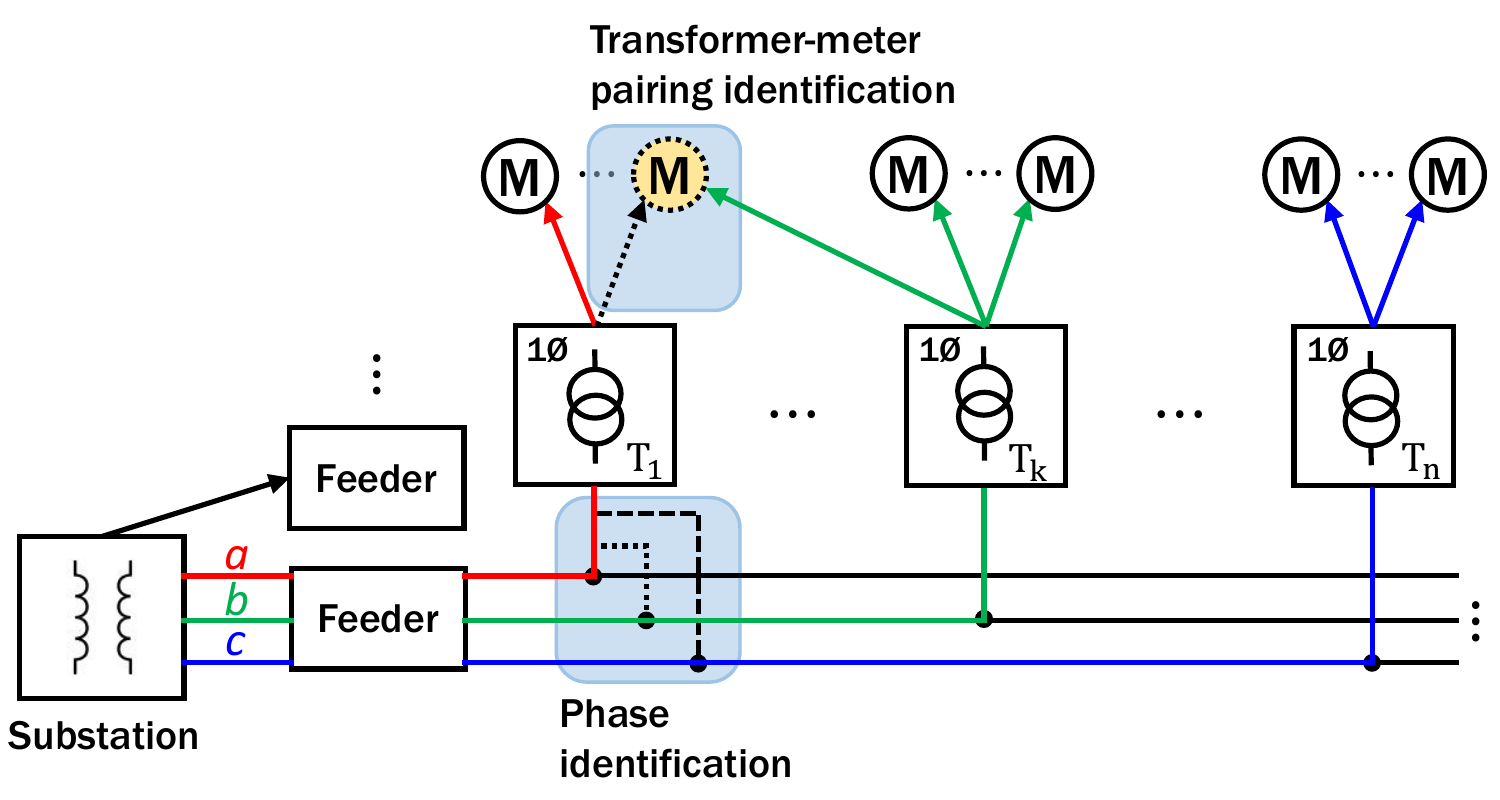}}
\vspace{-0.1in}
\caption{Illustration of a distribution feeder with 1-phase DTs supplying 1-phase customers. $a$, $b$, and $c$ are phase labels. $T_i$ represents the $i_{\mathrm{th}}$ DT. $M$ represents the customer meter.}
\label{fig1}
\end{figure}

\vspace{-0.1in}
\section{Methodology}\label{section2}
In this section, we provide an introduction to two fundamental problems in meter topology identification: meter phase and transformer-meter pairing. We then proceed to discuss the voltage correlation deterioration phenomena and elaborate on the utilization of PBDS as a means to mitigate this issue. Lastly, we present the PBDS based meter phase and transformer-meter pairing algorithms.

\vspace{-0.2in}
\subsection{Meter Topology Identification Problems}
\subsubsection{Problems Description}
In contemporary power distribution systems, the prevalence of two-phase loads has significantly diminished, primarily found in older residential installations and less common in new construction projects \cite{blalock2004first}. Therefore, this paper concentrates exclusively on the identification of phases for 1-phase loads. As illustrated in Fig.~\ref{fig1}, in a power distribution system serving multiple 1-phase circuits, the meter topology has two crucial pieces of information: the phase of the circuit and the distribution transformer supplying power to the meter. The phase information holds vital importance for utility engineers as it facilitates the assessment of 3-phase load balancing conditions across a distribution circuit. Similarly, transformer-meter pairing information enables utility engineers to estimate the transformer loading conditions.

\subsubsection{Data Requirement} \label{section2sub1}
The input data for the phase identification algorithm using voltage correlation-based approaches are presented in Table \ref{tab2}. Typically, voltage measurements taken at customer meters serve as the primary inputs for phase identification algorithms, as demonstrated in previous methods such as \cite{foggo2019improving, blakely2019spectral, blakely2020phase}. Additionally, in \cite{mitra2015voltage, wang2016phase}, voltage measurements at the feeder head can be incorporated to enhance the voltage correlation calculation. Conversely, for transformer-meter pairing identification, the input data typically consists of only smart meter data and the corresponding transformer labels assigned to each smart meter.

In comparison to existing methods, our approach achieves significant performance improvements by incorporating both power and voltage data. It is important to note that both types of data are readily available to utilities without the need for additional metering. This approach allows for the identification of voltage segments that display stronger correlations among meters sharing the same phase or supplied by the same transformer. Not utilizing both voltage and power measurements at the meter level not only results in an inefficient use of data but also contributes to a high false positive rate, as demonstrated in the results section.

\begin{table}[ht]
    \begin{center}
    \caption{Input Data Requirements for Phase Identification Problems.}
    \vspace{-.05in}
    \label{tab2}
    \resizebox{\linewidth}{!}{%
    \begin{tabularx}{\linewidth}{lcccccc}    
    \toprule
    \multirow{2}{*}{\textbf{Method}}  
    &\multicolumn{2}{c}{\textbf{AMI}} &   
    &{\textbf{Feeder}} 
    &\multirow{2}{2em}{\parbox{1\linewidth}{\centering \textbf{Phase} \\ \textbf{Label}}}       
    &\multirow{2}{2em}{\parbox{1\linewidth}{\centering \textbf{GIS} \\ \textbf{Data}}} \\ 
    \cmidrule{2-3} \cmidrule{5-5}
    &V (p.u.)   &P (\si{kW}) & &V (p.u.) \\
    \midrule
    \textbf{HCVC\cite{mitra2015voltage}}     &$\surd$   &$\times$ & &$\surd$   &$\surd$ &$\surd$    \\ 
    \textbf{PCA\cite{wang2016phase}}         &$\surd$   &$\times$ & &$\surd$   &$\surd$ &$\times$   \\
    \textbf{ITML\cite{foggo2019improving}}   &$\surd$   &$\times$ & &$\times$  &$\surd$ &$\times$   \\ 
    \textbf{SC\cite{blakely2019spectral}}    &$\surd$   &$\times$ & &$\times$  &$\surd$ &$\times$   \\ 
    \textbf{CAM-EC\cite{blakely2020phase}}   &$\surd$   &$\times$ & &$\times$  &$\times$ &$\times$  \\
    \textbf{Our Method}                      &$\surd$   &$\surd$  & &$\times$  &$\times$ &$\times$  \\ 
    \bottomrule
    \end{tabularx}}
    \end{center}
\end{table}

\subsubsection{Voltage Correlation Calculation} The underlying assumption of the voltage correlation based approach is that, when two 1-phase customers (indexed by $i$ and $j$) are on the same phase or supplied by the same 1-phase DT, the correlation between their voltage profiles, $V^i$ and $V^j$, are stronger than otherwise would have been. 

As will be explained in Section II.C, we calculate correlations between many short voltage segments (each within the range of a few hours) on two voltage profiles in order to determine the overall correlation between the two curves. This makes Pearson Correlation Coefficient (PCC) sufficient for computing correlations instead of using sophisticated non-linear correlation methods. PCC between $V^i$ and $V^j$, is calculated as
\begin{IEEEeqnarray}{lcr}
PCC(V^\mathrm{i}_\mathrm{n}, V^\mathrm{j}_\mathrm{n}) = \nonumber \\ \hfill \frac{\sum_{n=1}^{N}(V^\mathrm{i}_\mathrm{n}-\overline{V}^\mathrm{i})(V^\mathrm{j}_\mathrm{n}-\overline{V}^\mathrm{j})}{\sqrt{\sum_{n=1}^{N}(V^\mathrm{i}_\mathrm{n}-\overline{V}^\mathrm{i})^\mathrm{2}}{\sqrt{\sum_{n=1}^{N}(V^\mathrm{j}_\mathrm{n}-\overline{V}^\mathrm{j})^\mathrm{2}}}} \label{eq1}
\end{IEEEeqnarray}
where $\overline{V}$ is the average voltage, $n$ is the $n^{\mathrm{th}}$ data point, and $N$ is the number of points on the waveform.

Once the PCCs between meter pairs are calculated, we can either use a wide variety of classification methods to group meters together for phase group identification or compare the pair-wise PCCs to determine transformer-meter pairs. 

\subsection{Voltage Correlation Deterioration Phenomenon}   \label{section2sub2}
\subsubsection{Circuit Analysis}
\begin{figure}[t]
    \centering
  \subfloat[\label{2a}]{%
     \includegraphics[width=\linewidth, height=0.2\textheight]{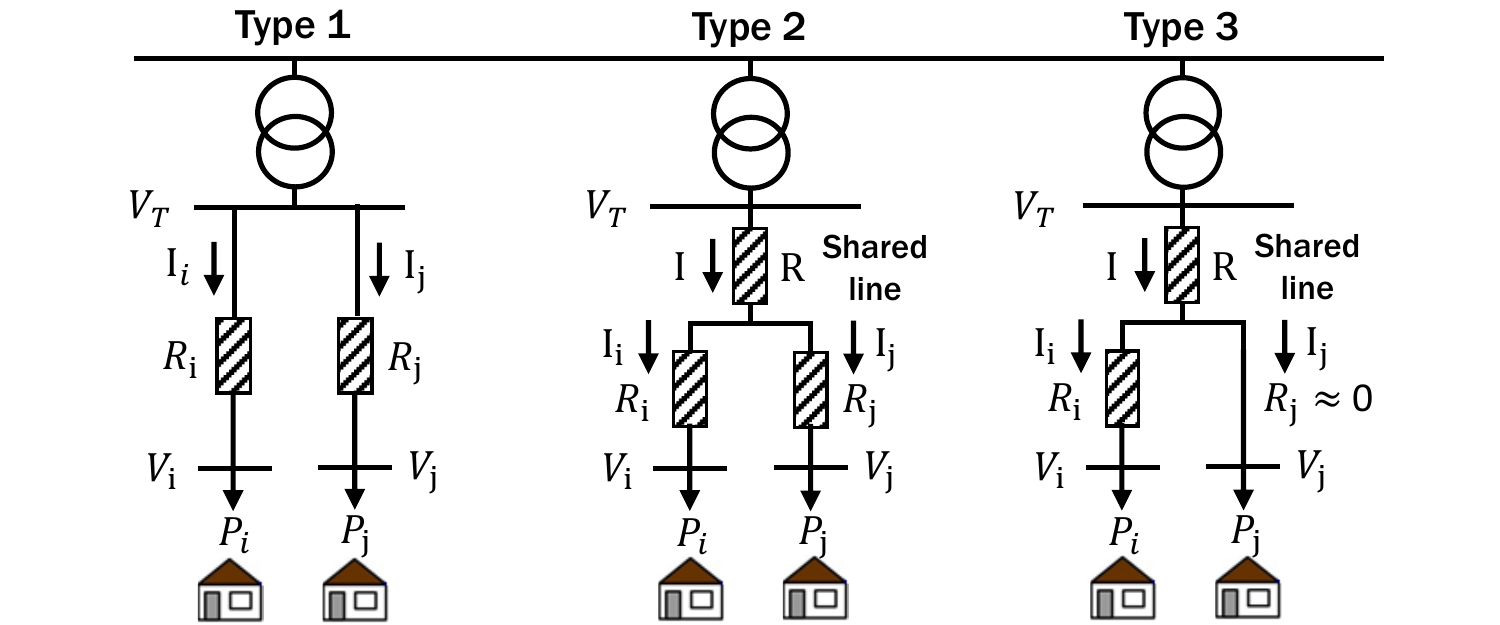}} \\ \vspace{-.1in}
  \subfloat[\label{2b}]{%
     \includegraphics[width=\linewidth, height=0.2\textheight]{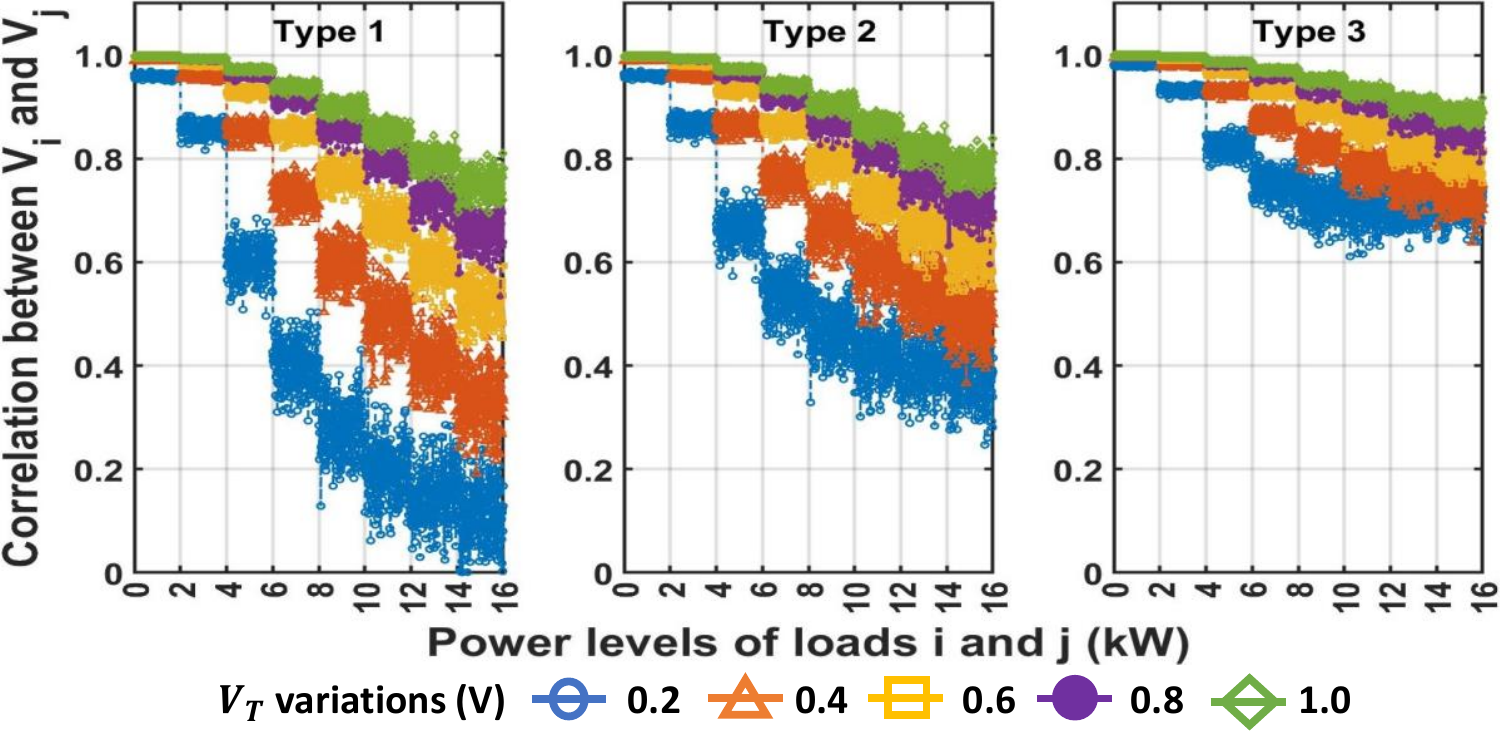}} \\ \vspace{-.1in}
  \subfloat[\label{2c}]{%
     \includegraphics[width=\linewidth, height=0.2\textheight]{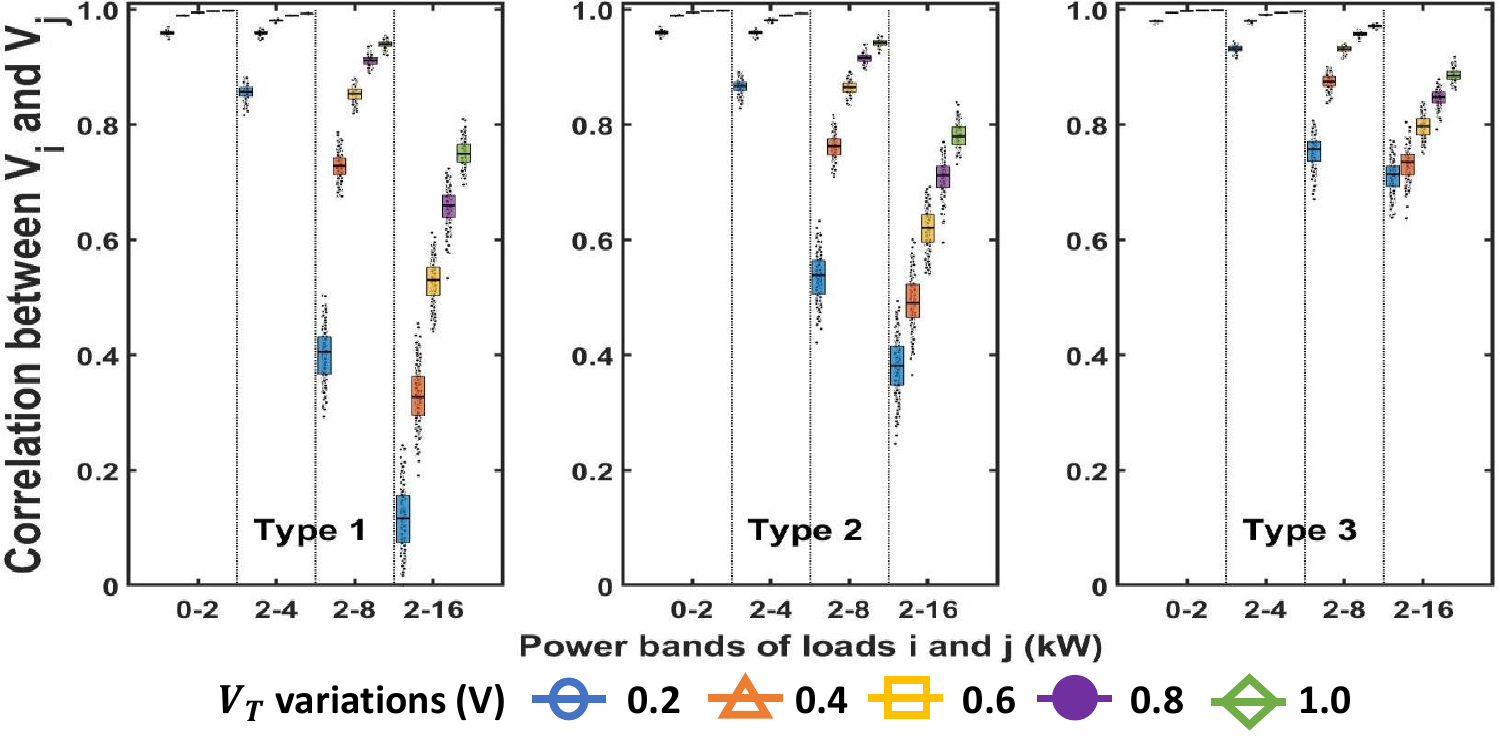}} \\ \vspace{-.07in}
\caption{(a) Three typical types of transformer secondary circuit connections, (b) PCC between $V_i$ and $V_j$ for types 1, 2 and 3 connections for different power consumption ranges, (c) Box plots of PCC between $V_i$ and $V_j$ for different power bands for types 1, 2 and 3 connections.}
\label{fig2}
\end{figure}

As shown in Fig.~\ref{fig2}(a), in relation to each other, two 1-phase loads connected to the same 1-phase transformer have three basic connection types: in-parallel, partially-parallel, and in-series. Denote $V_\mathrm{T}$, $I$, and $R$ as the DT voltage, shared line current and resistance, respectively. Denote $V_\mathrm{i}$ and $V_\mathrm{j}$, $I_\mathrm{i}$ and $I_\mathrm{j}$ as the system voltage and current of loads $i$ and $j$, respectively. Denote $R_\mathrm{i}$ and $R_\mathrm{j}$ as the resistance of the transformer secondary connection to loads $i$ and $j$. 

Then, we have
\vspace{-0.1in}
\begin{IEEEeqnarray}{rCl}
V_\mathrm{i} & = & V_\mathrm{T} - IR - I_\mathrm{i}R_\mathrm{i} \label{eq2} \\ 
V_\mathrm{j} & = & V_\mathrm{T} - IR - I_\mathrm{j}R_\mathrm{j} \label{eq3} \\
I & = & I_\mathrm{i} + I_\mathrm{j} \label{eq4}
\vspace{-0.1in}
\end{IEEEeqnarray}
Note that for type 1, $R=0$; for type 3, $R_\mathrm{j}=0$. 

From \eqref{eq2}--\eqref{eq4}, we have the following insights. 
\begin{itemize}
    \item When the transformer voltage is maintained close to its nominal values, the current on the secondary circuits (i.e., the lines connecting the transformer to the customer meters), $I_\mathrm{i}$ and  $I_\mathrm{j}$, are mainly determined by the power consumption levels of the $i^\mathrm{th}$ and $j^\mathrm{th}$ loads, $P_\mathrm{i}$ and $P_\mathrm{j}$. Therefore, if $P_\mathrm{i}$ and $P_\mathrm{j}$ are both low, then $I_\mathrm{i}$ and  $I_\mathrm{j}$ are low, causing a very small secondary voltage drop. Then, $V_\mathrm{i}$ and $V_\mathrm{j}$ will be very close to $V_\mathrm{T}$, making $V_\mathrm{i}$ and $V_\mathrm{j}$ highly correlated.
    \item However, when $P_\mathrm{i}$ and $P_\mathrm{j}$ are increasing, $I_\mathrm{i}$ and $I_\mathrm{j}$ will increase, leading to high secondary voltage drops on the secondary circuits. This will weaken the correlation between $V_\mathrm{i}$ and $V_\mathrm{j}$ significantly because the voltage variations are mainly determined by the values of $R$, $R_\mathrm{i}$, $R_\mathrm{j}$ and $|P_\mathrm{i}-P_\mathrm{j}|$.
\end{itemize}

To illustrate the phenomenon of voltage correlation deterioration with respect to the increase of local load consumption, Monte Carlo simulations are conducted to calculate PCCs between $V_\mathrm{i}$ and $V_\mathrm{j}$ for the three basic load connection types when the power levels of loads $i$ and $j$ vary randomly between 0 and 16 \si{\kW}; $V_\mathrm{T}$ varies randomly from 122 V within five bands (i.e., 0.2, 0.4, 0.6, 0.8, and 1 V); and $R$, $R_\mathrm{i}$ are the service drop resistance provided by the utility; $R_\mathrm{j}$ are set to 0.01 \si{\ohm}, 0.05 \si{\ohm}, and 0.05 \si{\ohm}, respectively. 

As shown in Fig.~\ref{fig2}(b), the voltage correlation deterioration for type 1 connection is the most prominent. When the output levels of both loads are above 6 kW while the variation of $V_\mathrm{T}$ is small (e.g., 0.2 V against 122 V), the correlation will drop below 0.4. The voltage correlation deterioration can exacerbate when the two loads are served by different DTs on the same phase due to the increase in line impedance. Therefore, excluding data sets that cause correlation deterioration is crucial for improving the accuracy of voltage-correlation based phase identification methods.

Next, we analyze the distribution of PCCs between loads $i$ and $j$ for low and high power bands under different transformer-level voltage fluctuations. Note that many 1-phase loads are residential. At night or daytime, when occupants are sleeping or not at home, the total electricity consumption is within a band of [0 2] \si{\kW}. For a load with no data segment below 2 \si{\kW}, we can widen the band. As shown in Fig.~\ref{fig2}(c), for voltage segments between low power bands (e.g., 0-2 or 2-4 kW), the secondary voltage drops are very small when $P_i$ and $P_j$ are low. This makes $V_i$ and $V_j$ nearly equal to $V_T$. As a result, the correlation between $V_i$ and $V_j$ is the strongest. 

However, when we increase the power band from 2 kW to 16 kW at a 2-kW step, the PCC value deteriorates, especially for type 1 connection. This is because, for loads with high electricity consumption levels, the secondary circuit voltage variations increase to a level equal to or even larger than the transformer-level voltage variations, making voltage correlations between loads with different power consumption patterns weaker. Therefore, we conclude that, for meter phase identification, selecting voltage segments within a low power band will increase identification accuracy by mitigating voltage correlation deterioration.

\begin{figure}[t]
    \centering
  \subfloat[\label{3a}]{%
     \includegraphics[width=\linewidth, height=0.12\textheight]{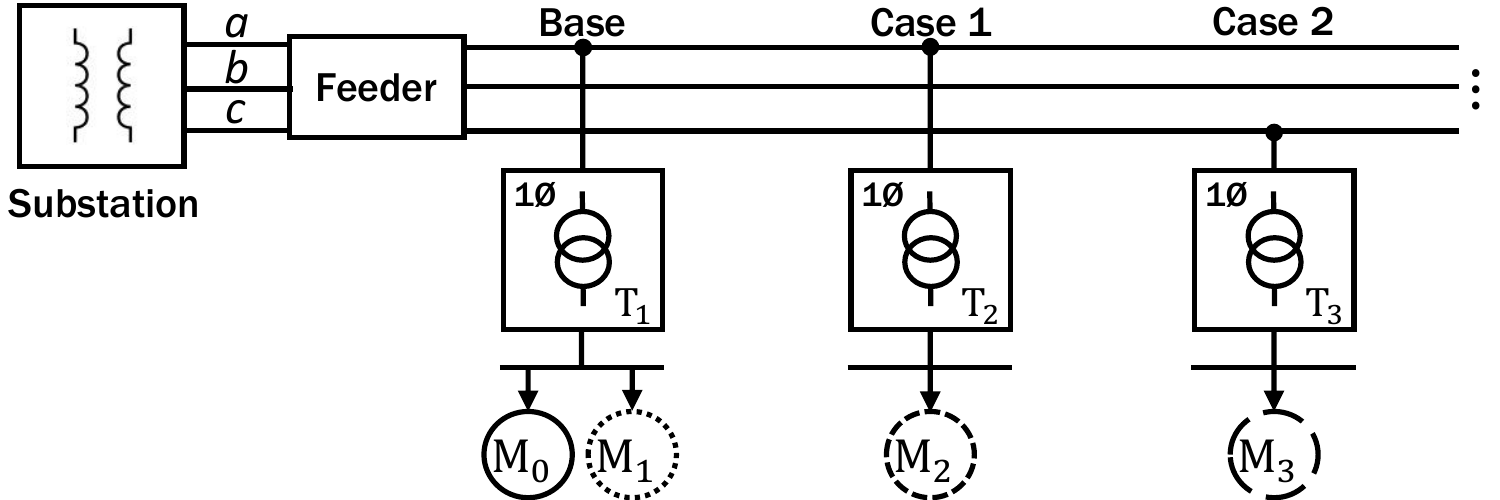}} \\ \vspace{-.1in}
  \subfloat[\label{3b}]{%
     \includegraphics[width=\linewidth, height=0.18\textheight]{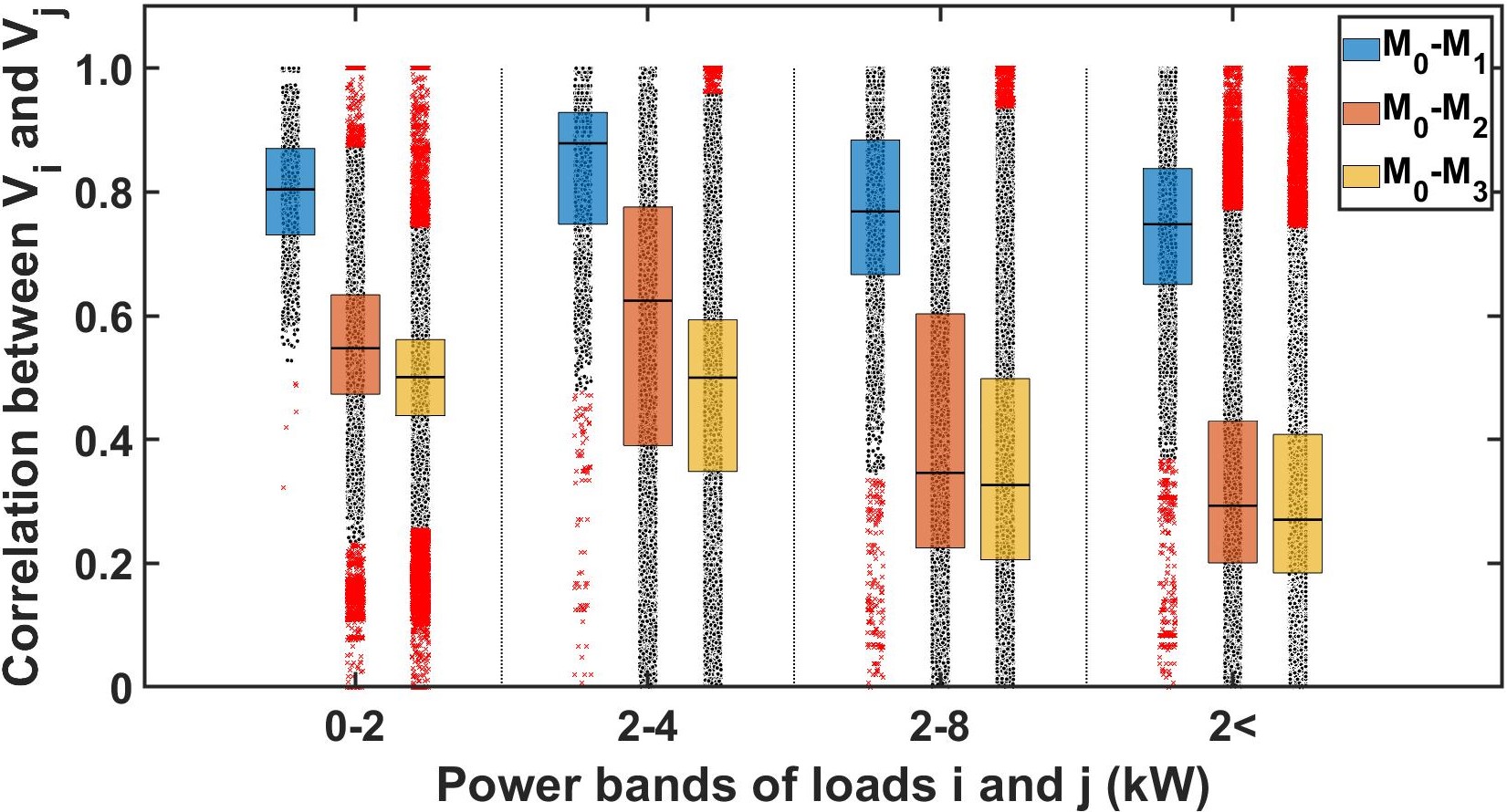}} \\ \vspace{-.1in}
\caption{(a) An illustration of meter locations, (b) Boxplots of PCCs between $V_i$ and $V_j$ for Feeder~12.}
\label{fig3}
\end{figure}

\subsubsection{Voltage Statistics from Actual Smart Meter Data}
To determine the topology relationship of a pair of meters, we need to know if they are served by the same 1-phase transformer ($M_0$-$M_1$), by different transformers on the same phase ($M_0$-$M_2$), or by different transformers on different phases ($M_0$-$M_3$). To further illustrate the voltage deterioration phenomena, the PCCs calculated using 4 power bands for three meter pairs on a real feeder are shown in Fig. \ref{fig3}, from which, we have the following observations:
\begin{itemize}
    \item Using a power band of [0 2] \si{\kW} or [2 4] \si{\kW} instead of higher power bands, the calculated PCCs of $M_0$-$M_1$ and $M_0$-$M_2$ are similar and significantly higher than the PCC of $M_0$-$M_3$. This shows for meter phase identification, using a low power band is preferred.
    \item Using a wide power band ($P>2$ \si{\kW}), the average PCC of $M_0$-$M_1$ shows a greater separation from the average PCCs of $M_0$-$M_2$ and $M_0$-$M_3$. This disparity arises due to the fact that, for high power consuming loads, the voltage fluctuations between loads supplied by different DTs are more significant compared to those between loads served by the same DT. Consequently, the voltage correlation becomes weaker. These findings highlight the preference for a wider power band when aiming to identify transformer-meter pairings. The validity of this observation is further supported by the results summarized in Table \ref{tab3}.
\end{itemize}

\vspace{-.1in}
\begin{table}[ht]
	\begin{center}
    \caption{Average PCC Difference by Power Band in Feeder~10.}
    \vspace{-0.1in}
    \label{tab3}
    \begin{tabular}{ccc}
    \toprule
        \multirow{2}{12em}{\parbox{1\linewidth}{\centering \textbf{Average PCC} \\ \textbf{difference between cases}}} & \multicolumn{2}{c}{\textbf{Power band [$P^\mathrm{-}$ $P^\mathrm{+}$]}} \\ 
        \cmidrule{2-3}
        & \textbf{0$\sim$2 \si{[\kW]}} & \textbf{2$<$ \si{[\kW]}} \\
	\midrule
	\textbf{$PCC_\mathrm{M_\mathrm{0}M_\mathrm{1}}-PCC_\mathrm{M_\mathrm{0}M_\mathrm{2}}$} & 0.2443 & \textbf{0.3859} \\
	\textbf{$PCC_\mathrm{M_\mathrm{0}M_\mathrm{1}}-PCC_\mathrm{M_\mathrm{0}M_\mathrm{3}}$} & 0.3046 & \textbf{0.4269} \\
	\textbf{$PCC_\mathrm{M_\mathrm{0}M_\mathrm{2}}-PCC_\mathrm{M_\mathrm{0}M_\mathrm{3}}$} & \textbf{0.0603} & 0.0411 \\
   \bottomrule 
   \end{tabular}
   \end{center}
\vspace{-0.3in}
\end{table}

\subsection{Power-Band based Data Segmentation (PBDS) Algorithm}
In \cite{pezeshki2012consumer, olivier2018phase}, the authors have proven that using voltage correlation for clustering is an effective method for customer phase identification. Based on the insight gained from the circuit analysis in Section~\ref{section2sub2}, we further demonstrate that using PBDS algorithm for extracting data segments can obtain data segments with stronger correlation between two time series $V_\mathrm{i}$ and $V_\mathrm{j}$. 

\begin{figure}[t]
\centerline{\includegraphics[width=\linewidth]{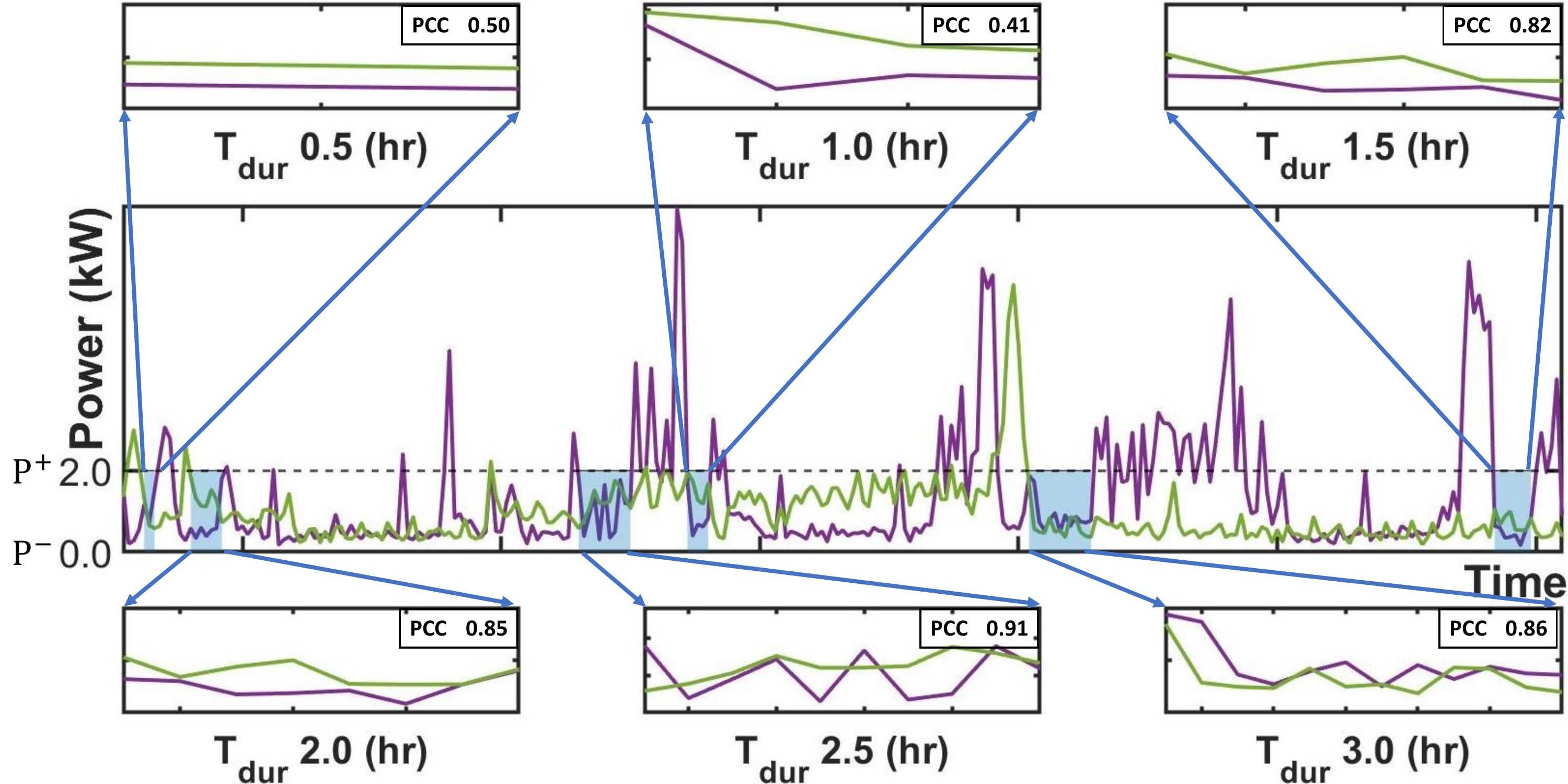}}
\caption{Demonstration of data segment selection with the power band [0 2] \si{\kW}, and $T_\mathrm{dur}$ between [0.5 3.0] \si{\hour} in 0.5 \si{\hour} increments in Feeder~10.}
\label{fig4}
\end{figure}

Define a power band as [$P^\mathrm{-}$ $P^\mathrm{+}$], where $P^\mathrm{-}$ and $P^\mathrm{+}$ are the power low and high limits, respectively. To ensure that the segments are sufficiently long to compute the PCC between two time series data sets, the minimum length, $T_\mathrm{dur}$, is used for excluding short data segments. Note that if less than two segments are found, the entire time series will be utilized for computing the voltage correlations.
Thus, the process for selecting data segments can be formulated as
\begin{IEEEeqnarray}{lCr}
P^\mathrm{-} \leq P_\mathrm{i,t} \leq P^\mathrm{+}, \quad T_\mathrm{dur} \leq m_\mathrm{i,k}\Delta T  \label{eq5} \\ 
P^\mathrm{-} \leq P_\mathrm{j,t} \leq P^\mathrm{+}, \quad T_\mathrm{dur} \leq m_\mathrm{j,k}\Delta T  \label{eq6} \\ 
PCC(V^\mathrm{M}_\mathrm{i}, V^\mathrm{M}_\mathrm{j}) = \nonumber \\ \hfill \frac{\sum_{k=1}^{K}(V^\mathrm{m_\mathrm{k}}_\mathrm{i}-\overline{V}^\mathrm{M}_\mathrm{i})(V^\mathrm{m_\mathrm{k}}_\mathrm{j}-\overline{V}^\mathrm{M}_\mathrm{j})}{\sqrt{\sum_{k=1}^{K}(V^\mathrm{m_\mathrm{k}}_\mathrm{i}-\overline{V}^\mathrm{M}_\mathrm{i})^\mathrm{2}}{\sqrt{\sum_{k=1}^{K}(V^\mathrm{m_\mathrm{k}}_\mathrm{j}-\overline{V}^\mathrm{M}_\mathrm{j})^\mathrm{2}}}} \label{eq7} \\ 
D(V^\mathrm{M}_\mathrm{i}, V^\mathrm{M}_\mathrm{j}) = 1 - \left| PCC(V^\mathrm{M}_\mathrm{i}, V^\mathrm{M}_\mathrm{j}) \right| \label{eq8}
\end{IEEEeqnarray}
where $i$ and $j$ are the indices of meter ($i,j \in \{1,\cdots,N_\mathrm{M}\}$), $t$ is the time steps ($t \in \{1,\cdots,T\}$), $\Delta T$ is the sampling interval, $\overline{V}^\mathrm{M}_\mathrm{i}$ and $\overline{V}^\mathrm{M}_\mathrm{j}$ are the mean values of $V^\mathrm{M}_\mathrm{i}$ and $V^\mathrm{M}_\mathrm{j}$, $k$ is the index of the data segment ($k \in \{1,\cdots,K\}$), $m_\mathrm{k}$ is the number of data points in the $k^\mathrm{th}$ data segments, $M = \{m_\mathrm{1},\cdots,m_\mathrm{K}\}$ is the set of the data segments. \eqref{eq7} computes the PCC matrix; \eqref{eq8} computes the correlation distance. 

\begin{table}[b]
	\begin{center}
		\caption{PCC Statistics of Data Segments by the [$P^\mathrm{-}$ $P^\mathrm{+}$] and $T_\mathrm{dur}$ for Feeder~10.}
		\vspace{-0.1in}
		\label{tab4}
		\centerline{\includegraphics[width=\linewidth]{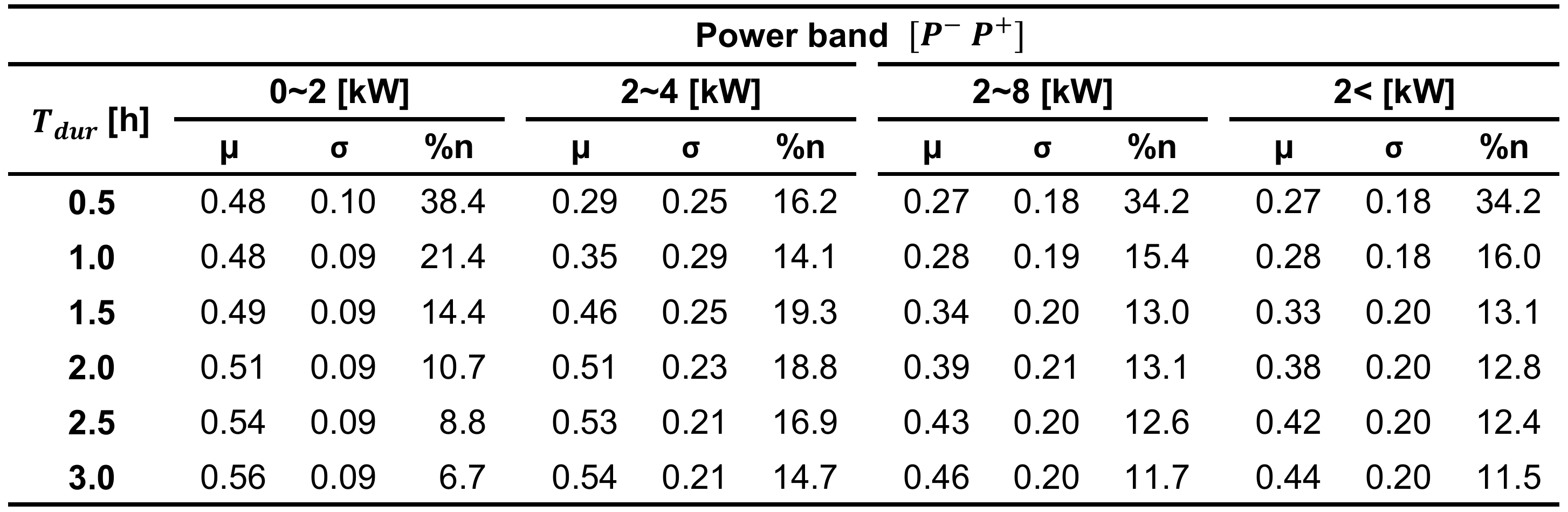}}  
	\end{center}
\vspace{-0.2in}
\end{table}

Note that \eqref{eq5} and \eqref{eq6} are the PBDS criteria (i.e., upper and lower power thresholds, and minimum duration). The data segment selection using low-power band ([0 2] kW) and various $T_\mathrm{dur}$ is shown in Fig.~\ref{fig4}. Only line segments between [$P^\mathrm{-}$ $P^\mathrm{+}$] with length longer than $T_\mathrm{dur}$ are selected. Table \ref{tab4} shows the PCC statistics of data segments that satisfy [$P^\mathrm{-}$ $P^\mathrm{+}$] and $T_\mathrm{dur}$. $\mu$, $\sigma$, and $\%n$ denote the mean, standard deviation, and share of total segments, respectively. It is essential to select an optimal parameter that reflects the characteristics of the feeder. The selection of the best [$P^\mathrm{-}$ $P^\mathrm{+}$] and $T_\mathrm{dur}$ values for phase identification and transformer-meter pairing will be discussed in Section~\rom{3}.

\begin{figure}[ht]
    \subfloat[\label{5a}]{%
        \includegraphics[width=.49\linewidth, height=0.16\textheight]{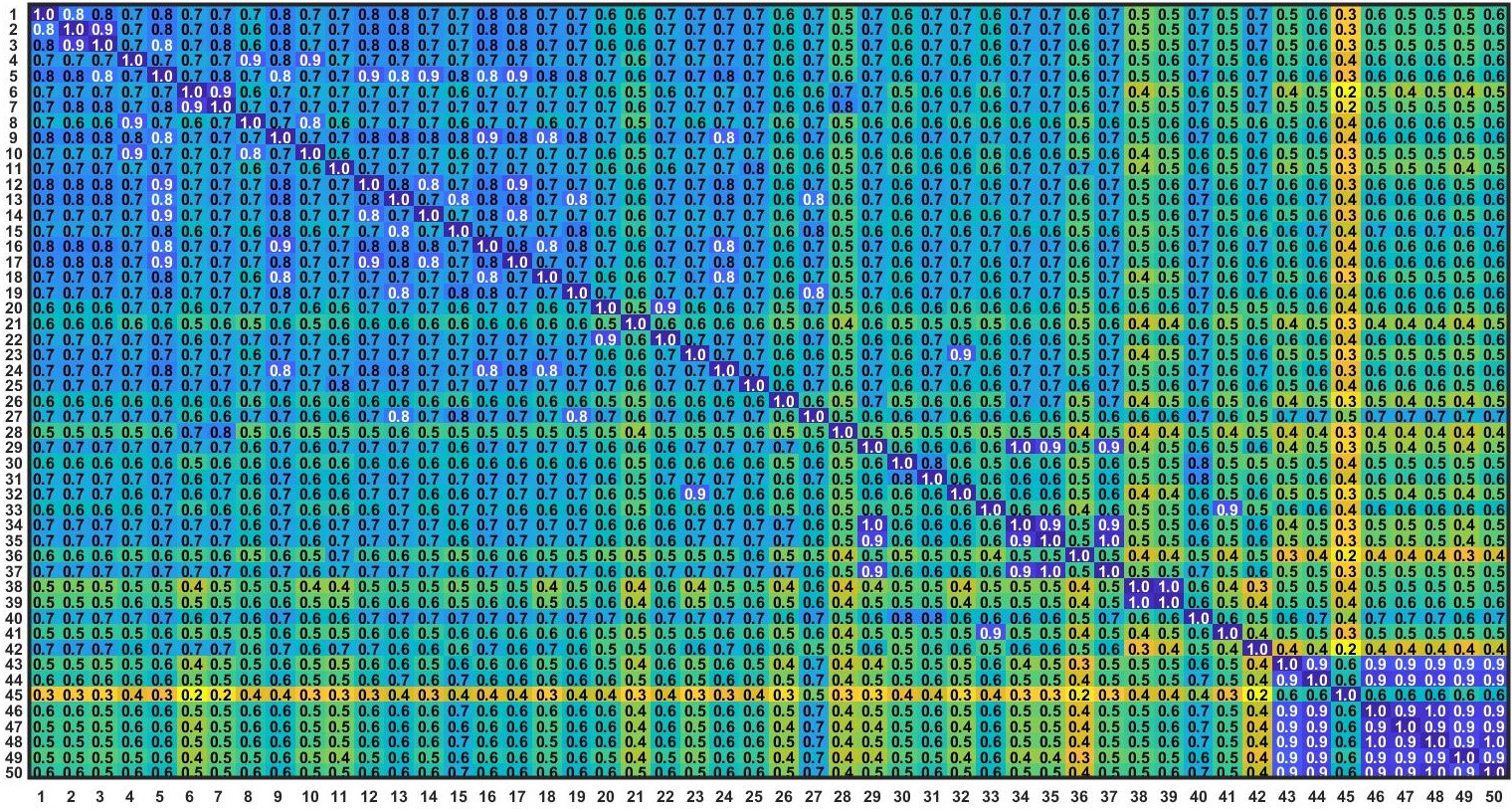}%
    }\hfill
    \subfloat[\label{5b}]{%
        \includegraphics[width=.49\linewidth, height=0.16\textheight]{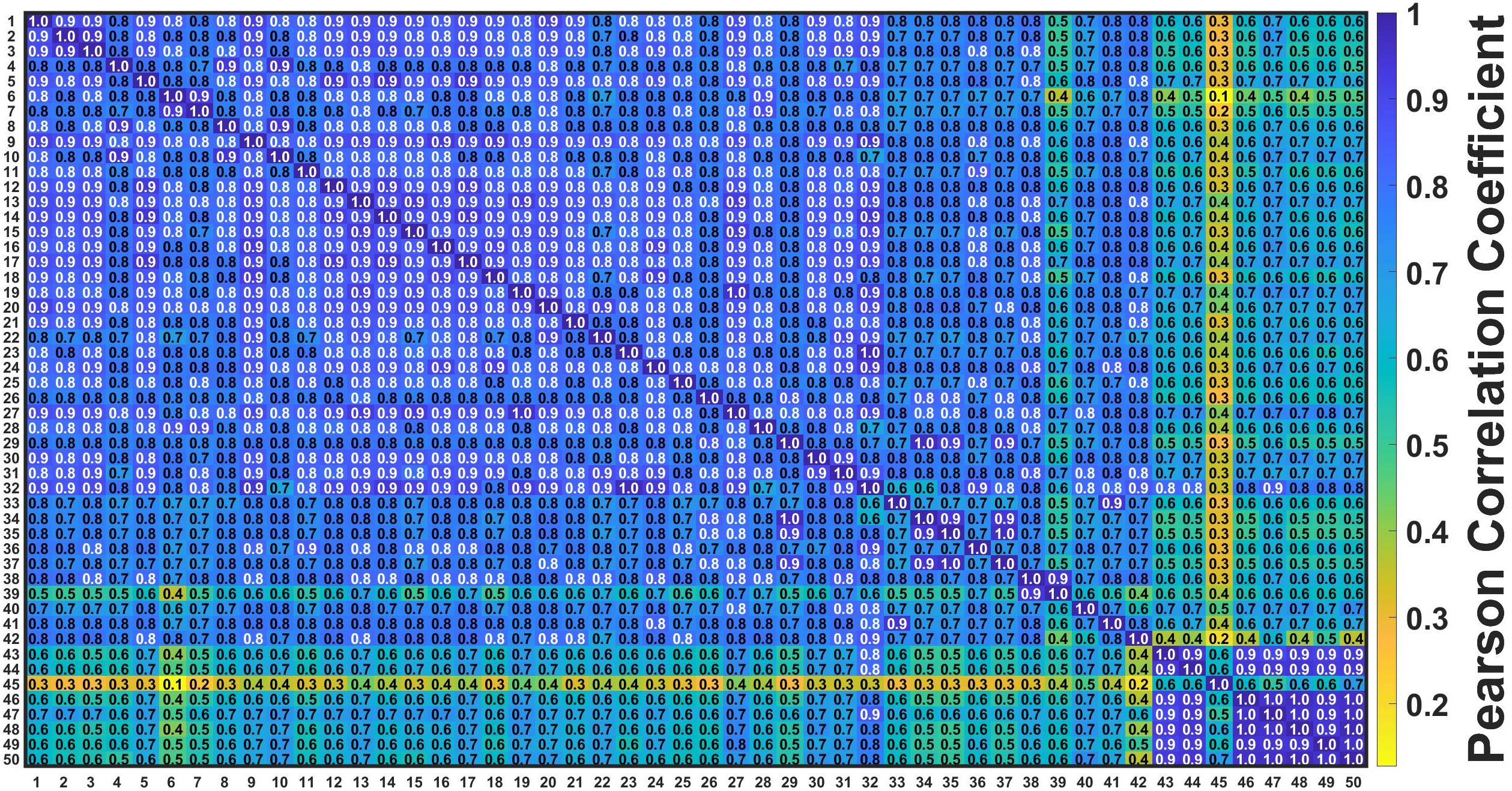}%
    }
    \vspace{-0.05in}
    \caption{Partial PCC matrix for Feeder 10, (a) without PBDS, (b) with PBDS ([$P^\mathrm{-}$ $P^\mathrm{+}$]=[0 1.3] and $T_\mathrm{dur}=2.5$).} 
    \label{fig5}
\vspace{-0.1in}
\end{figure}

As shown in Fig.~\ref{fig5}, when applying PBDS, PCCs calculated between meter pairs on the same phase show much stronger correlations than those on different phases, i.e., Fig.~\ref{fig5}(b) showing a much better contrast than Fig.~\ref{fig5}(a).

\vspace{-0.1in}

\section{Meter Topology Identification Algorithms}
In this section, we will present two PBDS based meter topology identification algorithms: phase identification and transformer-meter pairing.

\begin{figure}[b]
\centerline{\includegraphics[width=\linewidth]{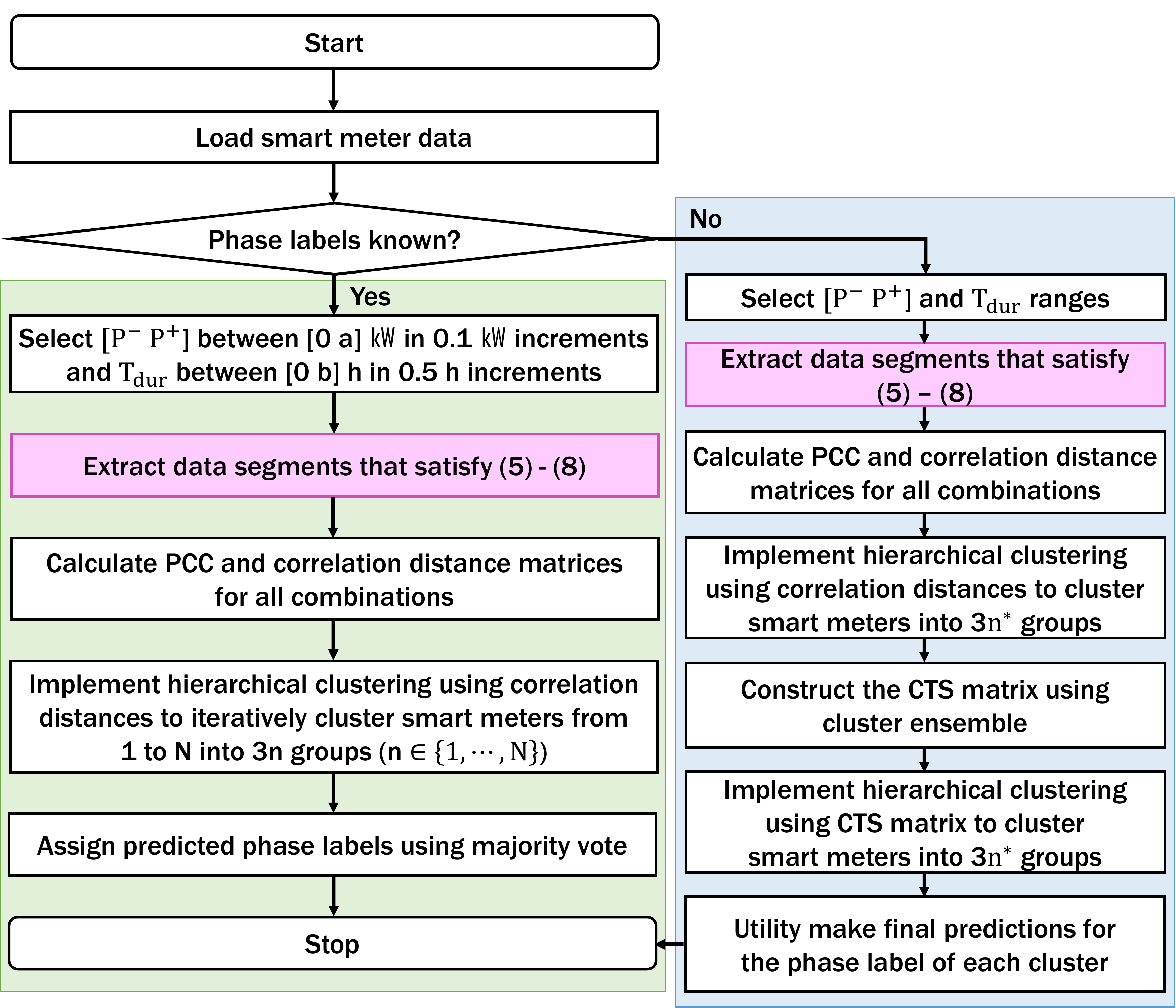}}
\vspace{-0.1in}
\caption{Flowchart of the PBDS based phase identification methodology.}
\label{fig6}
\end{figure}

\vspace{-0.1in}
\subsection{Phase Identification}
The key steps of the segmentation-based phase identification algorithm has been presented in our previous paper \cite{lee2022novel}. As shown in Fig. \ref{fig6}, the inputs of the algorithms are smart meter measurements including 15-\si{\minute} real power and voltage measurements. The PBDS steps are highlighted in the two shaded boxes. In this algorithm, we used the lowest power band, [0 2] kW, for extracting the voltage segments. This is because the correlation between meters on the same phase is the strongest within this band, as illustrated in Figs.~\ref{fig2} and \ref{fig3}. Then, the selected data segments are used for computing PCC and correlation distance between each pair of nodes. Based on correlation distances, the hierarchical clustering method will divide the smart meters into $3\times n$ clusters with $n$ increasing from 1 to $N$ (i.e., the number of clusters increases from 3 to 3$N$). If the phase labels are known, the majority vote mechanism presented in \cite{mitra2015voltage} will be used to assign phase labels to each cluster. If the phase labels are unknown, the CTS matrix constructed by the clustering ensemble method is proposed to determine the final clusters for utility engineers to label the phase for each cluster. Please refer to \cite{lee2022novel} for more details.

\vspace{-0.1in}
\subsection{Transformer-Meter Pairing Identification}
\begin{figure}[ht]
\centerline{\includegraphics[width=\linewidth, height=0.14\textheight]{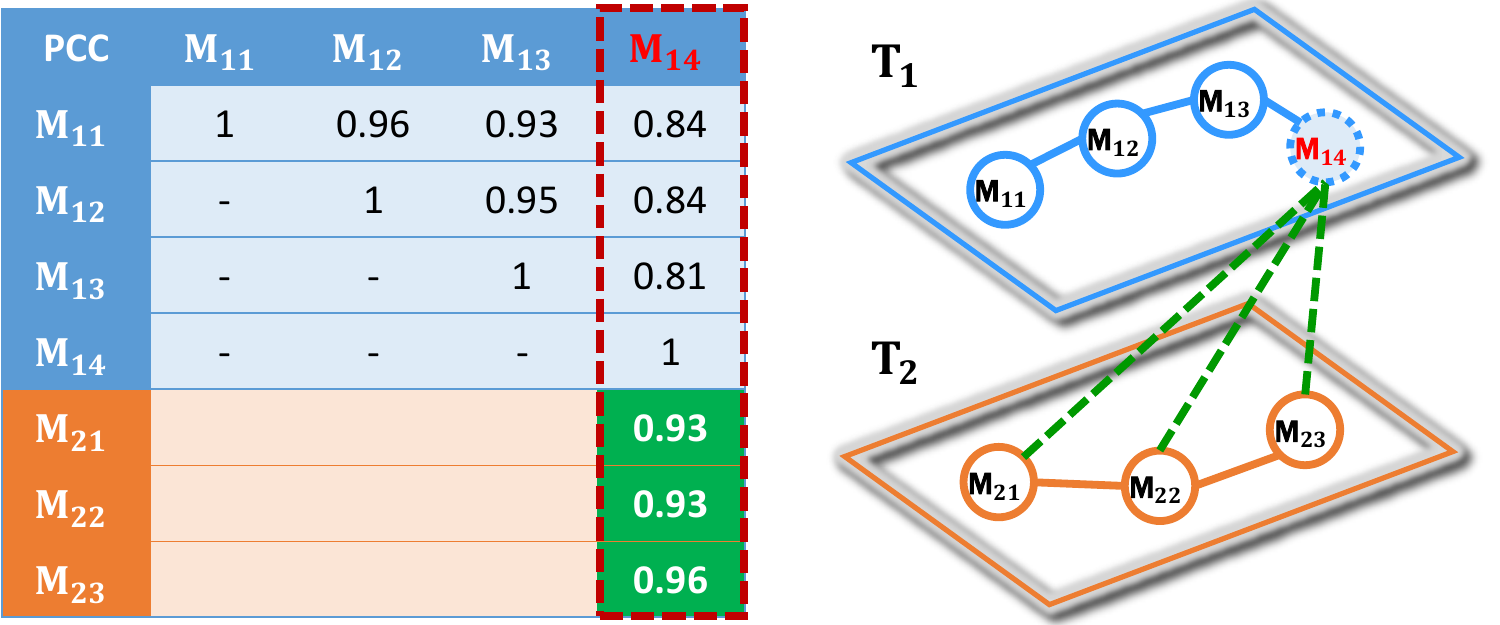}}
\vspace{-0.2in}
\subfloat[\label{7a}]{\hspace{.56\linewidth}}
\subfloat[\label{7b}]{\hspace{.44\linewidth}}   \vspace{-0.05in}
\caption{Detected error connection in Feeder~6. (a) PCC Matrix of transformers 1 and 2, (b) Connectivity map based on PCC.}
\label{fig7}
\end{figure}

When there are multiple meters connected to the same transformer, a meter may be strongly correlated with a couple of meters but not with all of them. Therefore, we can build a map based on the voltage PCC values to estimate their connections. As shown in Fig.~\ref{fig7}, $M_\mathrm{11} \sim M_\mathrm{14}$ belong to $T_\mathrm{1}$ and $M_\mathrm{21} \sim M_\mathrm{23}$ belong to $T_\mathrm{2}$. A connectivity map can be plotted by connecting the strongly correlated meters sequentially, starting with the meter with the maximum average voltage magnitude (e.g., $M_\mathrm{11}$, $M_\mathrm{21}$) under the same transformer. If $M_\mathrm{14}$ has a weaker correlation with other meters under $T_\mathrm{1}$ than with meters under $T_\mathrm{2}$, $M_\mathrm{14}$ and the two transformers will be flagged as having abnormal connections. 

\begin{figure}[t]
\centerline{\includegraphics[width=0.98\linewidth]{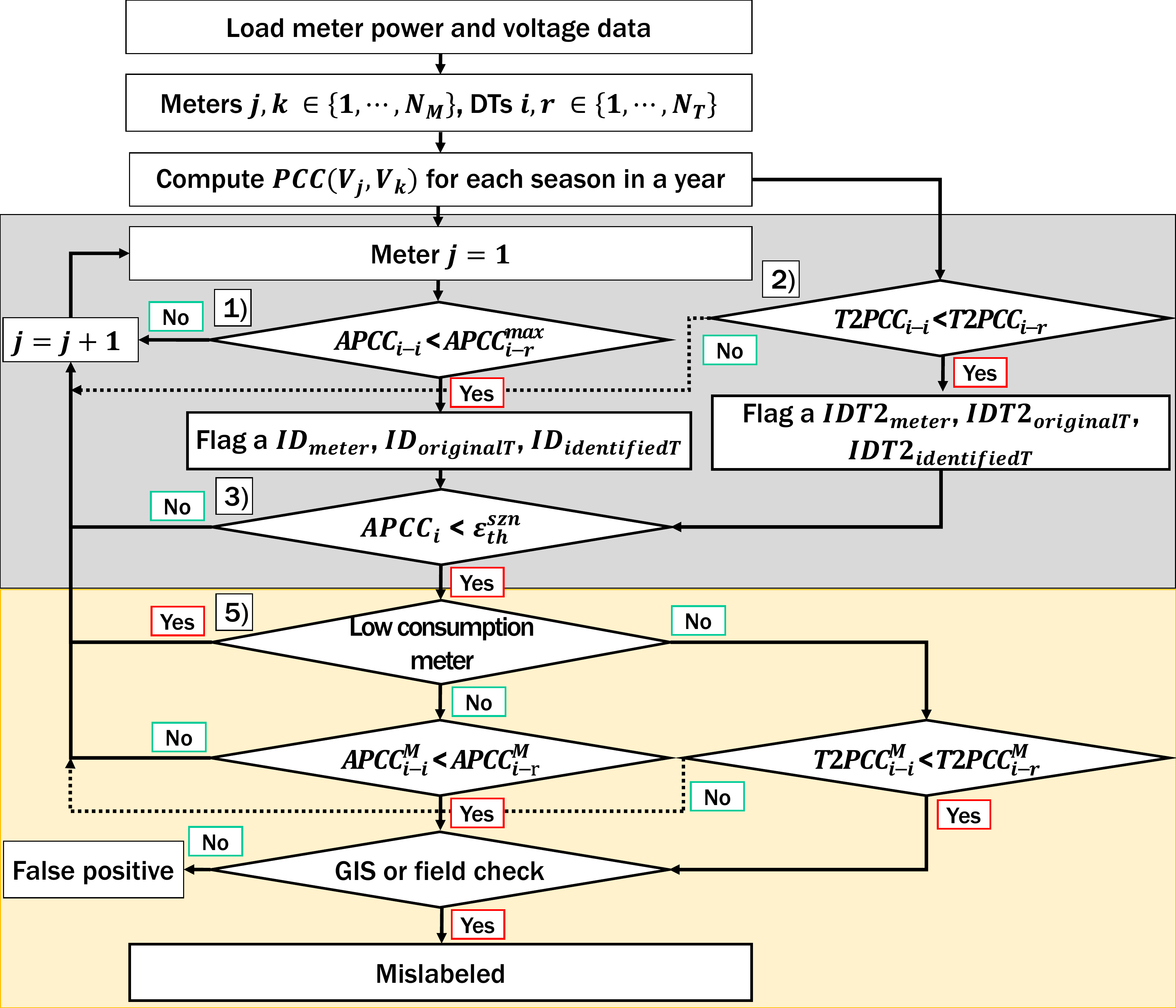}}
\caption{Flowchart of the PBDS based, two-stage transformer-meter pairing identification algorithm.}
\label{fig8}
\end{figure}

As shown in Fig.~\ref{fig8}, 15-minute smart meter data (including power and voltage measurements) are used to compute the PCC between meters for the same 1-phase transformer and with all other 1-phase transformers for each season in a year. We target the transformer serving more than one customer (i.e., connected to more than 1 meter). For the same transformer, based on PCC, a transformer-meter connection map can then be drawn. 

The transformer-meter pairing algorithm is a 2-stage process. In the first stage, the average PCC of a meter with respect to all the other meters under the same transformer is compared with the average PCC of the meter with all meters under another transformer to determine if the meter voltage is strongly correlated with its own meter group or another meter group. Because meters under the same transformer are not all strongly correlated, the average PCC of the top 2 correlated meters under the same transformer is compared with the average PCC of the top 2 meters under another transformer to identify mislabels that are not detected by the average PCC method. In both methods, if a transformer has a meter that is more strongly correlated with meters under another transformer, both transformers and their meters will be flagged for further inspection. Next, a seasonal PCC check is applied to identify moved meters. In the second stage, PBDS method is used to cross check to verify whether the meter is indeed mislabeled. 

After this, the location of all mislabelled meters will be identified on the google-map based geographical information system to verify their locations by utility engineers to determine if a field check is necessary. We will explain the flagging and pairing verification algorithm in the following subsections.

\subsubsection{Flag Abnormal Transformer-Meter Pairs using PCC Averages}
The average PCC for meter $j$ under transformer $i$ to all other meters under transformer $i$, $APCC_{\mathrm{i-i}}$, can be computed as
\vspace{-0.15in}
\begin{IEEEeqnarray}{lCr} \label{eq9}
 APCC_\mathrm{i-i}(j) = \frac{1}{N_{\mathrm{i}}-1} \sum_{k=1,k\neq j}^{N_{\mathrm{i}}}PCC(V_\mathrm{j},V_\mathrm{k}) 
\end{IEEEeqnarray}
where $N_{\mathrm{i}}$ is the number of meters supplied by transformer $i$.

The average PCC for meter $j$ under transformer $i$ to all other meters under transformer $r$, $APCC_{\mathrm{i-r}}$, can be computed as
\vspace{-.1in}
\begin{IEEEeqnarray}{lCr}\label{eq10}
 APCC_\mathrm{i-r}(j) = \frac{1}{N_{\mathrm{r}}} \sum_{k=1}^{N_{\mathrm{r}}}PCC(V_\mathrm{j},V_\mathrm{k})
\end{IEEEeqnarray}
where $N_{\mathrm{r}}$ is the number of meters supplied by transformer $r$.

Thus, if $APCC_{\mathrm{i-i}}(j)<APCC_{\mathrm{i-r}}(j)$, we will flag transformer $r$ as a possible host transformer for meter $j$ under transformer $i$. The IDs of the flagged meter, its current pairing transformer, the identified, possible pairing transformer will then be stored into three index vector, $ID_{\mathrm{meter}}$, $ID_\mathrm{originalT}$, and $ID_{\mathrm{identifiedT}}$. 

If there are more than one transformer satisfies the condition, the one with the biggest $APCC_{\mathrm{i-r}}(j)$ value will be selected. In our study, we observed that not all meters served by the same transformer are strongly correlated. Thus, using $APCC_{\mathrm{i-i}}(j)<APCC_{\mathrm{i-r}}(j)$ as the only criteria may skew the results if one or two meters are not strongly correlated with the other meters. 

\subsubsection{Flag Abnormal Transformer-Meter Pairs using the Average of top-2 PCCs}
To increase the identification accuracy, the weighted average of the top-2 PCC values, $T2PCC$, is calculated for meter $j$ of transformer $i$ using the PCCs from the top 2 correlated meters under transformer $i$ and under transformer $r$, respectively, as
\vspace{-0.1in}
\begin{IEEEeqnarray}{lCr}
 T2PCC_{\mathrm{i-i}}(j) = \sum_{k=1}^{2}w_\mathrm{k}PCC(V_\mathrm{j},V_\mathrm{k}) \label{eq11} \\
  T2PCC_{\mathrm{i-r}}(j) = \sum_{k=1}^{2}w_\mathrm{k}PCC(V_\mathrm{j},V_\mathrm{r}) \label{eq12}
\vspace{-0.05in}
\end{IEEEeqnarray}
where $w$ is the weighting factor. Thus, we can use 
$T2PCC_{\mathrm{i-i}}(j)<T2PCC_{\mathrm{i-r}}(j)$ to flag transformer $r$ as a possible host transformer for meter $j$ under transformer $i$. Similarly, the IDs of the meter, its current pairing transformer, the identified pairing transformer will be stored into index vectors, $IDT2_{\mathrm{meter}}$, $IDT2_\mathrm{originalT}$, and $IDT2_{\mathrm{identifiedT}}$ for further analysis.

\begin{figure}[b]
\centerline{\includegraphics[width=0.9\linewidth, height=0.13\textheight]{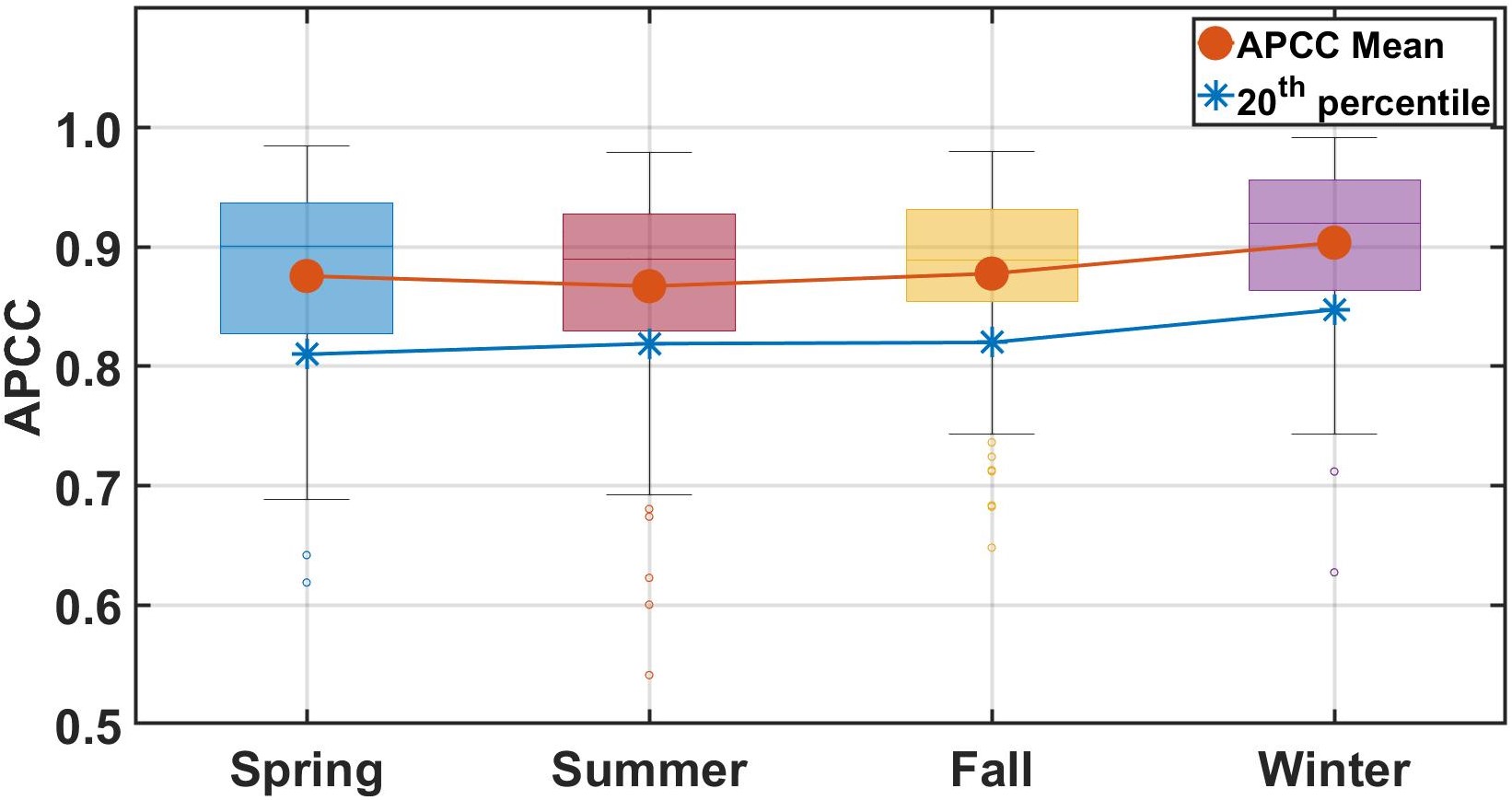}}
\vspace{-0.1in}
\caption{Seasonal PCC threshold ($\epsilon_\mathrm{th}^\mathrm{szn}$) for Feeder~6.}
\label{fig9}
\end{figure}

\subsubsection{Flag Abnormal Transformer-Meter Pairs using Seasonal PCC Variations}
Often times, the mislabelled transformer-meter pairing happens because a meter still in good condition has been moved from one transformer to another without its record being updated. Thus, when using the yearly data to compute $APCC$ or $T2PCC$, the values are much lower than what it should have been. This is because the PCC of the meter shows strong correlation with other meters under the labelled transformer for a period of time in a year, but after being moved away to another transformer, the correlation can drop drastically. Such transformer-meter pairing issues can be identified either by analyzing the time-series characteristic of $APCC$ or $T2PCC$ or by comparing the seasonal changes of $APCC$ or $T2PCC$.  

In this paper, to identify meters with lower than average PCC values, we compute the seasonal average PCC of all meters served by transformer $i$, $APCC_i$, as
\vspace{-0.05in}
\begin{IEEEeqnarray}{lCr}
APCC_i = \frac{1}{N_\mathrm{i}\times (N_\mathrm{i} - 1)}\sum_{j=1}^{N_\mathrm{i}}\sum_{k=1,k\neq j}^{N_\mathrm{i}}PCC(V_\mathrm{j},V_\mathrm{k}) \label{eq13} 
\vspace{-0.05in}
\end{IEEEeqnarray}
Note that the seasonal PCC threshold ($\epsilon_{\mathrm{th}}^{\mathrm{szn}}$) is selected to detect sudden correlation drops. In this paper, based on the PCC distribution shown in Fig.~\ref{fig9}, we set $\epsilon_{\mathrm{th}}^{\mathrm{szn}}$ to be $20^{\mathrm{th}}$ percentile so that all meters with $APCC_i \leq \epsilon_{\mathrm{th}}^{\mathrm{szn}}$ are flagged. In practice,  $\epsilon_{\mathrm{th}}^{\mathrm{szn}}$ can be selected a little higher or lower depending on the cost of field verification.

\subsubsection{Identify the Possible Transformer(s) Supplying the Flagged Meter}
It is possible that for a mislabeled transformer-meter pair, we identify more than one transformer that may supply the possibly mislabelled meter, especially if we are using different criterion listed above. However, we find that it is sufficient to select the most likely transformers to be the one with the highest $APCC$ and $T2PCC$ values calculated by \eqref{eq9} and \eqref{eq11}, and then confirm the selection by using seasonal $APCC$. In theory, we can conduct exhaustive search for all transformers that satisfy $APCC_{\mathrm{i-i}}(j)<APCC_{\mathrm{i-r}}(j)$ or $T2PCC_{\mathrm{i-i}}(j)<T2PCC_{\mathrm{i-r}}(j)$. However, in practice, this approach is often times unnecessary because the 2-stage checking will further verify the mislabelled pair and the subsequent field inspection can quickly reveal the actual transformer-meter pairing information. Thus, it is crucial to identify the mislabeled transformer-meter pairs and less crucial to use the algorithm for identifying what transformer actually supplies the mislabelled meter.

\subsubsection{PBDS based Second Stage Verification}
The goal of the 2-stage verification is to eliminate false positives. As shown by the correlation analysis in Section~\ref{section2sub2}, a meter with very lower power consumption tends to have high correlation with another meter on the same phase even though the two meters are supplied by different transformers.  Because in such cases, the local power variations are very small, which cannot lead to sufficiently large $APCC$ gaps.  

Therefore, in the second stage verification process, we first exclude meters with very low power consumption when computing for $APCC$ and $T2PCC$ using \eqref{eq9}, \eqref{eq10}, \eqref{eq11}, and \eqref{eq12}. This can be achieved by using the data segments that satisfy $P>P^{\mathrm{TH}}_{\mathrm{LP}}$, where $P^{\mathrm{TH}}_{\mathrm{LP}}=1$ \si{\kW} is the low power threshold. Again, to guarantee the length of the data segment is sufficient for computing PCC, we set the $T_\mathrm{dur}>1$ \si{\hour}. 

Thus, for a meter, if there is no data segment that can satisfy $P>1$ \si{\kW}, it will be marked as a "low consumption meter". For a flagged meters, if there is no data segment satisfies this condition, we will unflag the meter. If after the low power segments are removed, for a flagged meter, we have $APCC_{\mathrm{i-i}}^{\mathrm{M}}(j)>APCC_{\mathrm{i-r}}^{\mathrm{M}}(j)$ or $T2PCC_{\mathrm{i-i}}^{\mathrm{M}}(j)>T2PCC_{\mathrm{i-r}}^{\mathrm{M}}(j)$, the meter and the corresponding transformer will be unflagged. The remaining transformer-meter pairs will go through a GIS check and subsequent field verification.

\begin{table}[b]
\vspace{-0.05in}
	\begin{center}
		\caption{Parameter Selection for Real Feeder Data Sets.}
		\vspace{-0.1in}
		\label{tab5}
		\centerline{\includegraphics[width=\linewidth]{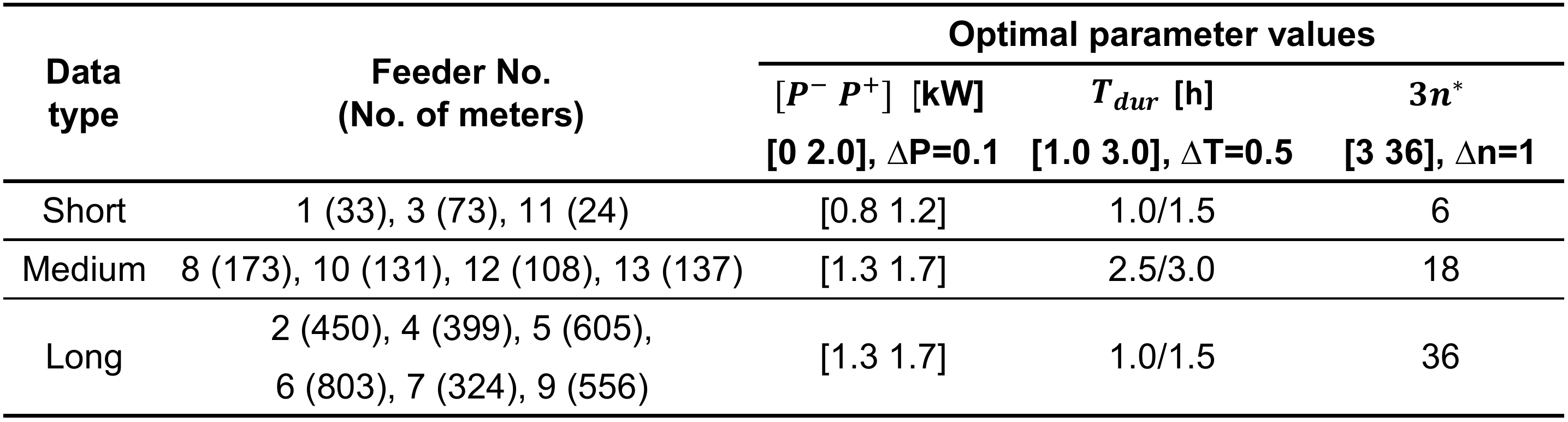}}    
	\end{center}
\vspace{-0.15in}
\end{table}

\begin{table*}[t]
	\begin{center}
		\caption{Case~1: Phase Identification Results and Performance Comparison.}   \vspace{-0.1in}
		\label{tab6}
		\centerline{\includegraphics[width=\linewidth, height=0.36\textheight]{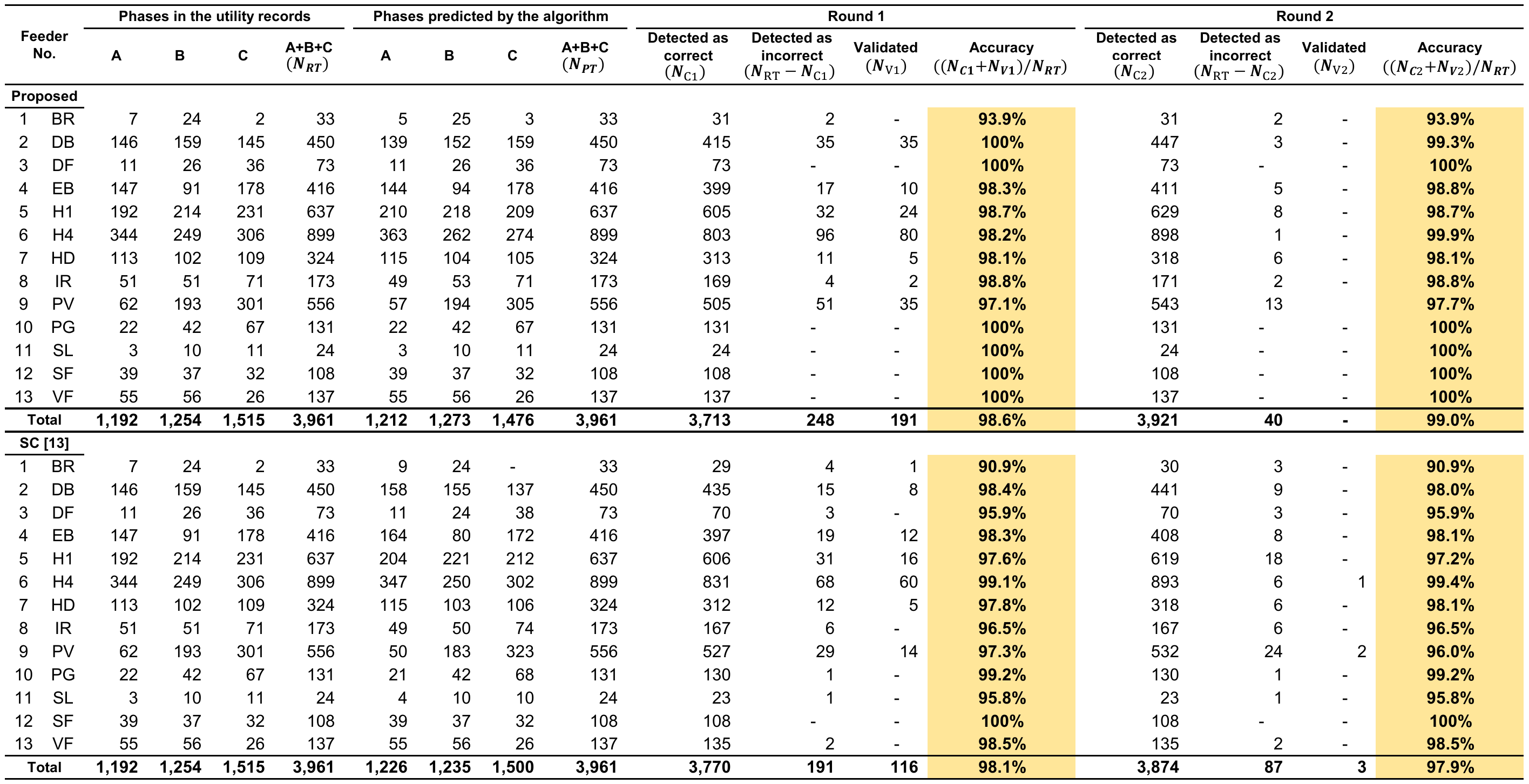}}    
	\end{center}
\vspace{-0.4in}
\end{table*}

\vspace{-0.1in}
\section{Simulation Results}
One year of 15-minute smart meter data collected from 3,961 customers on 13 real feeders in the North Carolina area are used to test the performance of the two developed algorithms. The utility also provided the phase label for each 1-phase transformer. Within this dataset, voltage segments that satisfy the criteria of a power band ranging between [0 2] kW and a minimum duration of 1 hour or more were identified in 99.9\% of cases. If there were less than two voltage segments in the ensemble, the full time-series voltage profiles will be used to calculate the voltage correlations between the two meters. All tests were carried out on a desktop computer equipped with an Intel Core i7-7700@3.6 GHz processor and 16 GB of memory.

\begin{figure}[t]
\centerline{\includegraphics[width=\linewidth, height=0.18\textheight]{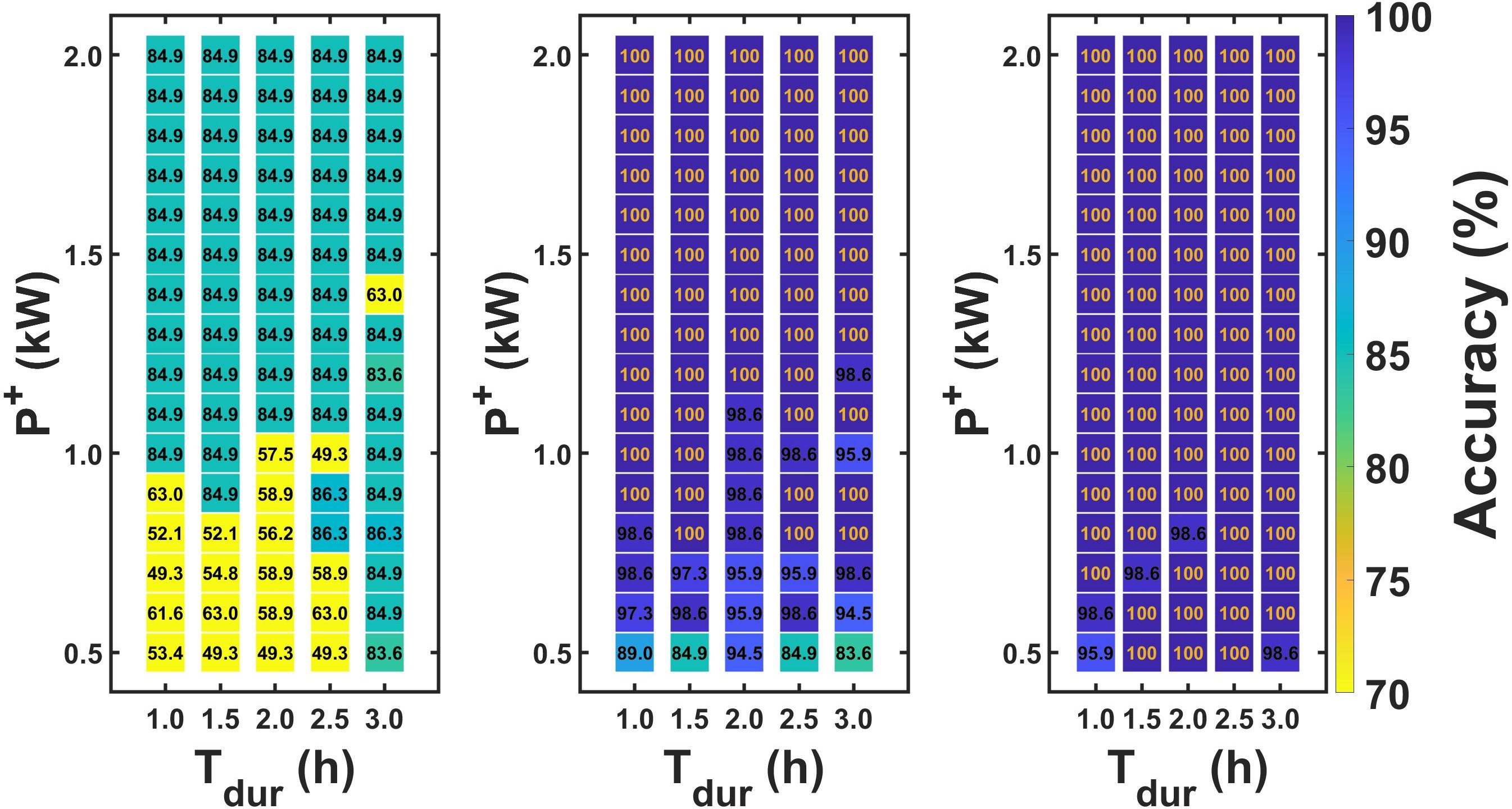}}   \vspace{-0.15in}
\subfloat[3 clusters \label{10a}]{\hspace{.38\linewidth}}
\subfloat[6 clusters \label{10b}]{\hspace{.24\linewidth}}
\subfloat[9 clusters \label{10c}]{\hspace{.34\linewidth}}
\caption{Phase identification accuracy for three different numbers of clusters with varying parameters in Feeder~3. The accuracy of most parameter combinations converged when $3n^*=6$.}
\label{fig10}
\end{figure}

\vspace{-0.15in}
\subsection{Case 1: Phase Identification \textbf{with} Phase Labels}
To select the optimal parameters for each feeder, we conduct Monte Carlo simulations for the range of parameters shown in Table \ref{tab5}. If [$P^\mathrm{-}$ $P^\mathrm{+}$] and $T_\mathrm{dur}$ are too small, there will be very few data segments satisfying the low-power and minimum data length conditions. This renders the correlation calculation unreliable. As shown in Table \ref{tab5}, the maximum accuracy occurs when $P^\mathrm{-} \geq 0.8$ \si{\kW} and $T_\mathrm{dur}\geq 1.0$ \si{\hour}. Therefore, parameter ranges [$P^\mathrm{-}$ $P^\mathrm{+}$] between [0 0.4] \si{\kW} and $T_\mathrm{dur}$ between [0 0.5] \si{\hour} are removed from the subsequent simulations. The average computation times for parameter tuning on short, medium, and long feeders are 2.9 \si{\hour}, 3.6 \si{\hour}, and 39.5 \si{\hour}, respectively.

The findings indicate that the optimal number of clusters varies depending on the number of meters present on a circuit. As shown in Fig.\ref{fig10}, by maintaining $P^\mathrm{-}=0$ and increasing $P^\mathrm{+}$ from 0.5 to 2, using 6 clusters yields the highest identification accuracy, with no further improvement achieved by increasing the number of clusters to 9. Analysis reveals that short feeders with tens of customers require only 6 clusters, medium feeders with a few hundred customers require 12 clusters, and long feeders with over 400 customers require 36 clusters. The findings suggest that circuits with longer laterals can benefit from grouping customers into more clusters. This is due to the potential weakening of voltage correlation between two meters on the same phase but supplied by different lateral circuits, as discussed in the type I circuit analysis in Section\ref{section2sub2}. Consequently, dividing meters into additional groups enhances clustering accuracy. 

\begin{table*}[t]
	\begin{center}
		\caption{Case~2: Phase Identification Results and Performance Comparison.}
		\vspace{-0.1in}
		\label{tab7}
		\centerline{\includegraphics[width=\linewidth, height=0.36\textheight]{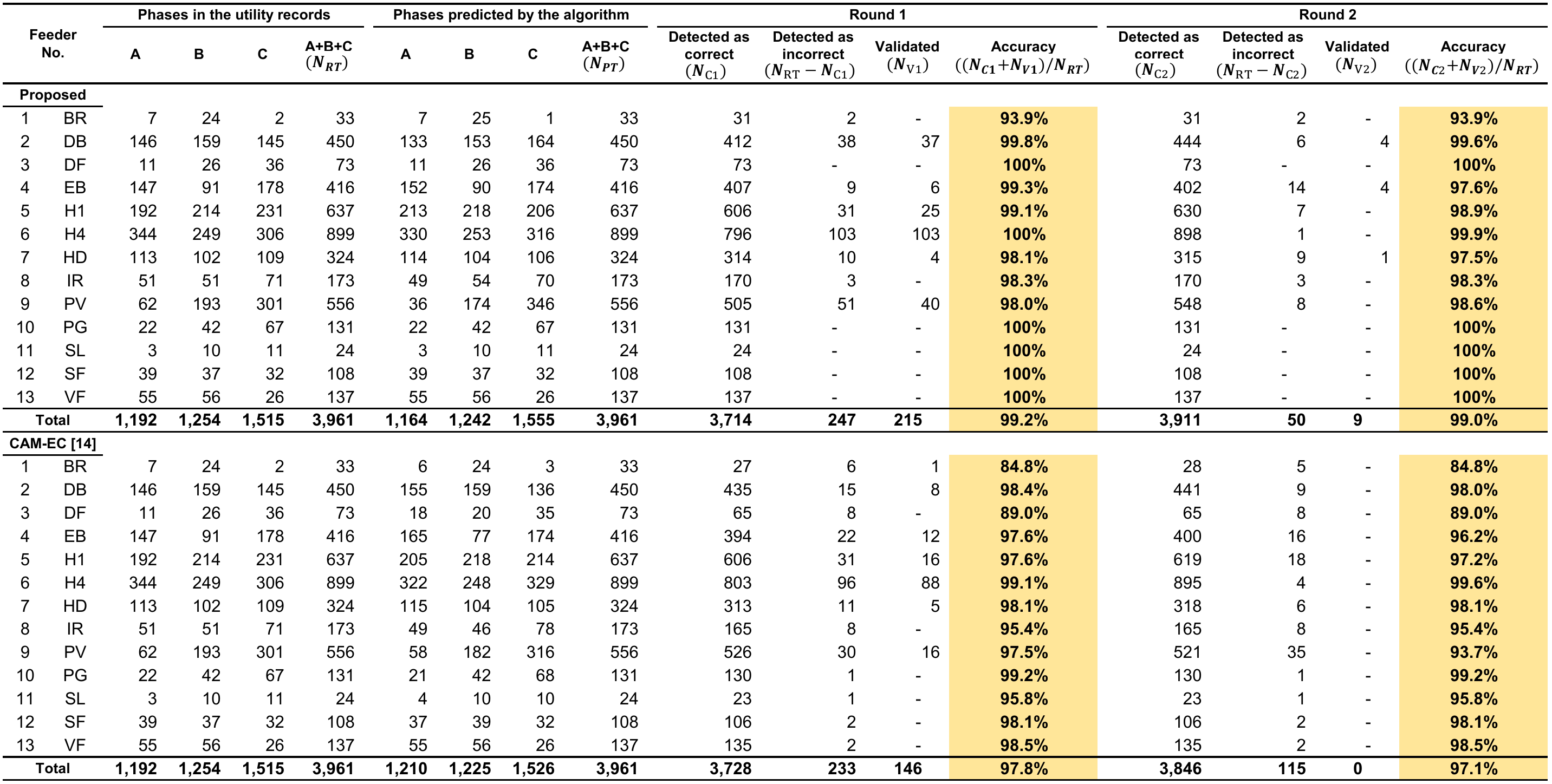}}    
	\end{center}
\vspace{-0.3in}
\end{table*}

\begin{table*}[ht]
	\begin{center}
		\caption{Transformer-Meter Pairing Identification Results (ID: Average PCC; IDT2: Top-2 PCC).}
		\vspace{-0.1in}
		\label{tab8}
		\centerline{\includegraphics[width=\linewidth, height=0.18\textheight]{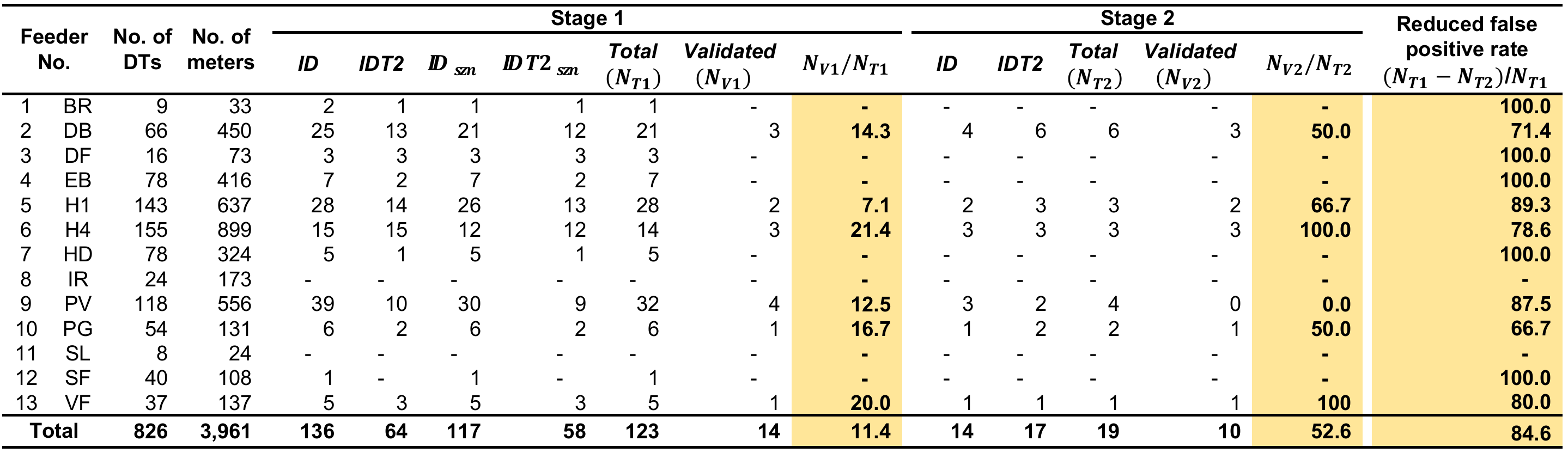}}
	\end{center}
\vspace{-0.4in}
\end{table*}

We validate our algorithm performance through a two-round field verification process. In the first round, any mislabeled meters were corrected. Subsequently, we reran the algorithm to identify any remaining mislabeled meters. If such meters were found, utility engineers conducted a second field check to determine the accuracy of the identified mislabeled meters. This process will provide more thorough evaluation of the performance achieved in the first round. In practice, it is essential to establish a threshold that allows for the occurrence of a small number of "false negatives" rather than striving to eliminate all "false positives." This is particularly important when utility engineers prioritize the identification of mislabeled meters for ensuring reliable operation.

As shown in Table \ref{tab6}, the proposed approach consistently outperformed the Spectral Clustering (SC) method \cite{blakely2019spectral} across all data sets. The window size for running SC was set to 4 days, and the number of clusters for each data set was determined as specified in Table \ref{tab5}. This outcome demonstrates that using selected data segments, our algorithm captures voltage correlation between meters sharing the same phase better. In round~1, our method accurately predicts 93.7\% of the original utility phase labels, with only 1.4\% of "mislabeled" data found to be "false positive" following site verification.

The "false positives" meters are approximately 1\% of the total meters, the majority of which are situated at the feeder's far end and connected to the service transformer via lengthy secondary lines. These secondary lines exhibit higher impedance (as discussed in Section~\ref{section2sub2}), resulting in weaker voltage correlations between these meters and others supplied by the same transformer. This weakened correlation is the main reason for misclassification. Additional, a few "false positive" meters are characterized by exceptionally high power consumption, in which cases, only a few voltage segments are used for PCC calculation, leading to large variances and causing misclassification.

\vspace{-0.15in}
\subsection{Case 2: Phase Identification \textbf{without} Phase Labels}
The performance of the proposed method is compared with the Co-association Matrix Ensemble Clustering (CAM-EC) introduced in \cite{blakely2020phase}. In CAM-EC, the window size is fixed at 4 days, and the number of clusters is determined as follows: 3 and 6 clusters for short feeders, 6, 12, and 18 clusters for medium feeders, and 6, 12, 15, and 30 clusters for long feeders. 

As shown in Table~\ref{tab7}, the PBDS method outperformed CAM-EC in all cases. The identification accuracy of the PBDS method reached 100\% for nearly half of the cases, whereas CAM-EC typically achieved accuracy below 100\%. This difference in performance can be attributed to the fact that CAM-EC generates a cluster ensemble by combining different numbers of clusters, which can lead to inaccurate results unless the optimal number of clusters is determined for each specific data set. We also observed from the results that typically, feeders with a larger number of meters tend to have higher identification accuracy. This is because meters on longer feeders tend to display more distinct voltage profiles on different phases. 

In the second round, although we can still identify some mislabeled meters, there is an sharp increase in the "false positive" rate. This observation suggests that in certain locations along a feeder, relying solely on APCC values may not be sufficient for achieving accurate phase identification, as discussed in the previous section.

\vspace{-0.1in}
\subsection{Results of Pairing Identification}
Table \ref{tab8} reports the results of transformer-meter pairing identification. In stage~1, the numbers of flagged abnormal transformer-meter pairs are represented by $ID$ and $IDT2$, which are calculated using the average PCC values from \eqref{eq9} and \eqref{eq10}, and the top-2 average PCC values from \eqref{eq11} and \eqref{eq12}, respectively. $ID_{\mathrm{szn}}$ and $IDT2_{\mathrm{szn}}$ indicate the number of meters excluded when checking $APCC_{\mathrm{i}}>\epsilon_{\mathrm{th}}^{\mathrm{szn}}$. In stage~2, $ID$ and $IDT2$ represent the number of abnormal transformer-meter pairs after removing false positives using the PBDS method. Subsequently, utility engineers performed a GIS check and field verification on the flagged meters from both stages to identify potential false positives and false negatives. For Feeder 6, which has the highest number of DTs and meters, the computational time required was 75 minutes. This demonstrates the good scalability of the proposed algorithm, as it can handle larger systems effectively without any issues.

Overall, the 2-stage pairing identification algorithm achieves 52.6\% accuracy and a false positive rate of 47.4\%. In stage 1, using the PCC average (ID) and top-2 PCC average (IDT2) methods without PBDS, approximately 3.4\% (136 out of 3,961 meters) and 1.6\% (64 out of 3,961 meters) are labeled as "abnormal" transformer-meter pairs, respectively. Applying the seasonal PCC check reduces the numbers to 117 (14\% reduction) and 58 (9.4\% reduction), respectively. PBDS's impact on identification accuracy is evident in stage 2, as the subsequent method reduces false positives from 117 to 14 (88.0\% reduction) and from 58 to 17 (70.7\% reduction), respectively. This minimizes the need for field checks, alleviating utility concerns about high false positive rates.

Feeder 9 experiences multiple reconfiguration, resulting in the highest number of false positives and the lowest detection effectiveness ($N_{\mathrm{V2}}/N_{\mathrm{T2}}$) among all the other feeders (refer to Table \ref{tab8}).
Additionally, the five remaining false positives are meters located at the far ends of distribution transformers (DTs), leading to weak voltage correlation with other meters on the same transformer.

\vspace{-0.1in}
\section{Conclusion}
In this paper, we applied power-band-based data segmentation (PBDS) in two meter topology identification algorithms: customer phase identification and transformer-meter pairing identification. In customer phase identification, we utilized the low power band condition to select voltage segments. In transformer-meter pairing identification, we employed a two-stage method where in the second stage, PBDS is used for reducing false positives by excluding meters with extremely low power consumption and obtaining voltage segments with larger correlation gaps. This step played a crucial role in distinguishing between meters supplied by the same transformer and those supplied by different transformers on the same phase. Through simulation on thirteen real feeders, our proposed algorithm outperformed existing approaches, significantly improving the accuracy of meter topology identification without introducing significant computational complexity.
\vspace{-0.1in}

\ifCLASSOPTIONcaptionsoff
\newpage
\fi 
\bibliographystyle{IEEEtran}
\bibliography{IEEEabrv,MyRefs}

\end{document}